\documentclass[referee,a4paper,12pt,traditabstract]{swsc} 

\usepackage{graphicx}
\usepackage{txfonts}
\usepackage{subfigure}
\usepackage{epstopdf}

\usepackage{amsmath}
\usepackage[displaymath,mathlines]{lineno}
\usepackage[authoryear,round]{natbib}
\usepackage[backref]{hyperref}
\usepackage{url}

\hypersetup{colorlinks=true,citecolor=cyan,urlcolor=cyan,linkcolor=blue}

\begin{document}

   \title{Ensemble Forecasting of Major Solar Flares: Methods for Combining Models}

   
   \titlerunning{Ensemble Forecasting of Solar Flares}

   \authorrunning{Guerra {\it et. al.}}

   \author{J. A. Guerra\inst{1}, S. A. Murray\inst{2,3}, D. S. Bloomfield\inst{4} \and P. T. Gallagher\inst{3,2}
          }

   \institute{Physics Department, Villanova University
              800 E Lancaster Ave. Villanova, PA 19085, USA\\ \email{\href{mailto:jordan.guerraaguilera@villanova.edu}{jordan.guerraaguilera@villanova.edu}}\thanks{Corresponding author}
            \and
             School of Physics, Trinity College Dublin, Ireland
                      \and
             School of Cosmic Physics, Dublin Institute for Advanced Studies, Ireland
           \and
             Department of Mathematics, Physics and Electrical Engineering, Northumbria University, Newcastle upon Tyne, NE1 8ST, UK
             }


 
  \abstract
   {One essential component of operational space weather forecasting is the prediction of solar flares. With a multitude of flare forecasting methods now available online it is still unclear which of these methods performs best, and none are substantially better than climatological forecasts. Space weather researchers are increasingly looking towards methods used by the terrestrial weather community to improve current forecasting techniques. Ensemble forecasting has been used in numerical weather prediction for many years as a way to combine different predictions in order to obtain a more accurate result. Here we construct ensemble forecasts for major solar flares by linearly combining the full-disk probabilistic forecasts from a group of operational forecasting methods (ASAP, ASSA, MAG4, MOSWOC, NOAA, and MCSTAT). Forecasts from each method are weighted by a factor that accounts for the method's ability to predict previous events, and several performance metrics (both probabilistic and categorical) are considered. It is found that most ensembles achieve a better skill metric (between 5\% and 15\%) than any of the members alone. Moreover, over 90\% of ensembles perform better (as measured by forecast attributes) than a simple equal-weights average. Finally, ensemble uncertainties are highly dependent on the internal metric being optimized and they are estimated to be less than 20\% for probabilities greater than 0.2. This simple multi-model, linear ensemble technique can provide operational space weather centres with the basis for constructing a versatile ensemble forecasting system -- an improved starting point to their forecasts that can be tailored to different end-user needs.
   }

   \keywords{Solar flares forecasting --
                Ensembles --
                Weighted linear combination
               }

   \maketitle

\section{Introduction}

\begin{figure}[!t]
\centering
\includegraphics[width=17pc]{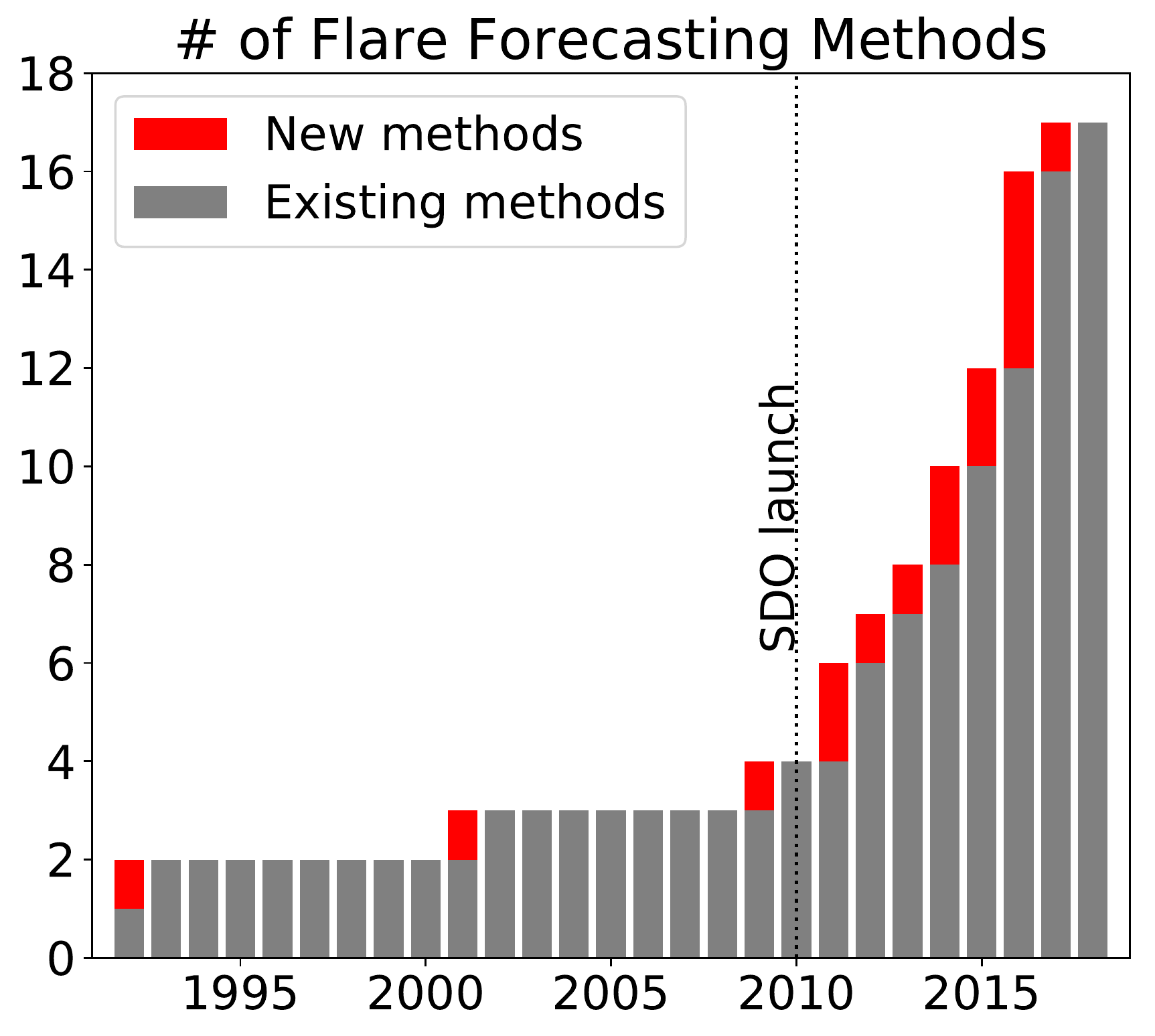}
\includegraphics[width=19pc]{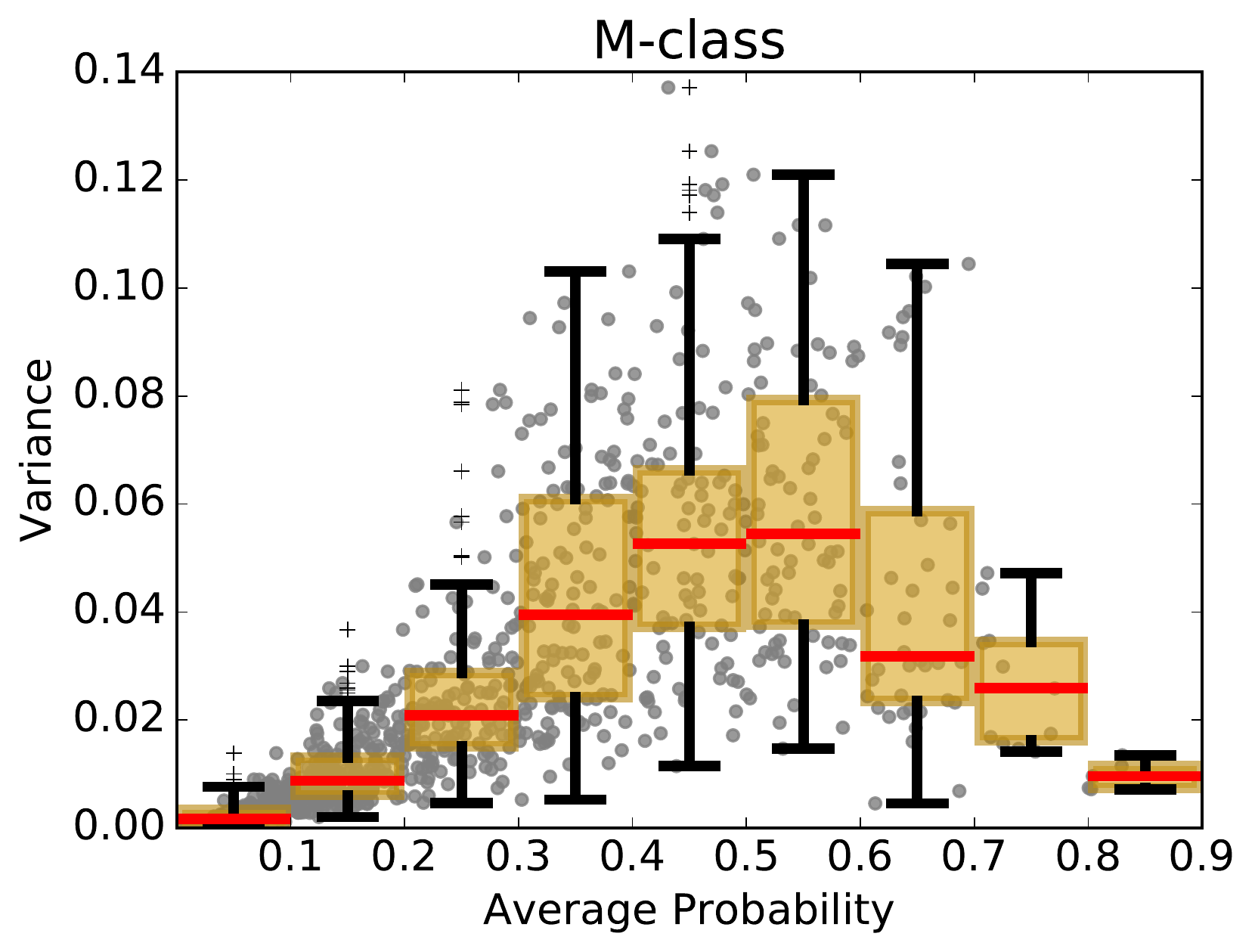}
\caption{{\it Left:} Number of flare forecasting methods publicly available per year since 1992. For each year, existing methods (grey) and new methods (red) are displayed. Since 2010 the number of flare forecasting methods has increased at an average of approximately three every two years. This information was partially gathered from \cite{Leka2019}, the NASA/GSFC Community Coordinated Modeling Center (CCMC) archive of forecasts, and other operational centre online archives. The earliest date when the first forecast was made available in these sources was used for the purposes of this figure. {\it Right:} forecast variance vs. average forecast for a six-method group of probabilistic forecasts for M-class flares between 2015 and 2017. Variance is lower when the average forecast is closer to zero or one.}
\label{fig:spread_forecasts}
\end{figure}

Predicting when a solar flare may occur is perhaps one of the most challenging tasks in space weather forecasting due to the intrinsic nature of the phenomenon itself (magnetic energy storage by turbulent shear flows + unknown triggering mechanism + magnetic reconnection), the lack of more appropriate remote-sensing data, and the rarity of the events, particularly for large (i.e., X-class) flares \citep{Leka2018,Hudson2007}. Yet the need for more accurate, time-sensitive, user-specific, and versatile forecasts remains relevant as the technological, societal, and economical impact of these events becomes more evident with time \citep{Tsagouri2013}. In the past decade the number of flare forecasting methods has increased rapidly at an annual average rate of $\sim$ 1.5
(Fig.~\ref{fig:spread_forecasts}, left panel). Much of this accelerated growth seems possible thanks to the availability of new science mission data from {\it Solar Dynamics Observatory} \citep[SDO;][]{Pesnell2012}, which provides high-quality solar imagery with an operational-like routine.

Differences in input data, training sets, empirical and/or statistical models used among different forecasting methods make it difficult to directly compare performances across all methods \citep{Barnes2016}. Just by looking at the probabilistic forecasts for M-class flares from a group of six different methods (see Section~\ref{Section:Data}) during three years (2015--2017) reveals that the variance of the probabilities around the average value is significantly larger away from 0 and 1 (Fig.~\ref{fig:spread_forecasts}, right panel). That is, for a particular time, if solar conditions are such that there is only a low chance of observing an M-class flare, all methods report similar low chances. Similarly, if solar conditions favor a high chance of observing the flare, all methods seem to report similar high chances. However, if there is only a moderate chance of observing a flare, forecasts in this case can range from low to very high. 

In an operational environment, space weather forecasters are often faced with the responsibility of issuing alerts and making decisions based on forecasts like those described above. However, in cases where different methods provide very different forecasts, it can be difficult to know which method can be more accurate given the specific solar conditions. It is in these cases that using the different forecasts to create a combined, more-accurate prediction may be advantageous. Ensemble forecasting, although successfully used for terrestrial weather practices for decades, is fairly new in space weather \citep{Knipp2016,Murray2018}. In the field of flare forecasting, \citet{Guerra2015} demonstrated the applicability of multi-model input ensemble forecasting for the flare occurrence within a particular active region. Using a small statistical sample ( only four forecasting methods and 13 full-passage active-region hourly forecast time series) the authors showed that linearly combining probabilistic forecasts, using combination weights based on the performance history of each method, makes more accurate forecasts. In addition, \citet{Guerra2015} suggested that combining forecasts which are intrinsically different (i.e.,  automatic/software versus human-influenced/experts) have the potential to improve the prediction in comparison to the case in which only forecasts of similar type are used. However, the small data sample used in such analysis (events and forecasts) is not statistically significant for fully quantifying how much ensembles can improve prediction of flares.

In this study, the ensemble forecasting method presented in \citet{Guerra2015} is expanded to include more forecasting methods and a larger data sample, with a particular focus on analyzing full-disk forecasts that are used more widely by operational centers. The effects of considering different performance metrics and linear combination schemes are modelled and tested. Section~\ref{Section:Data} presents and briefly describes the data sample employed (forecasts, forecasting methods used to create these, and observed events). In Section~\ref{Section:Models} the ensemble models are described. Main results are presented and discussed in Section~\ref{Section:Results}. Discussion is organized around the constructed ensembles, comparison among them, and a brief demonstration of uncertainty analysis is also included. Finally, conclusions and potential future work are outlined in Section~\ref{Section:Conclusions}.

\section{Forecast Data Sample}
\label{Section:Data}
In this investigation, full-disk probabilistic forecast time series for the occurrence of M- and X- class flares were used from six different operational methods. Table~\ref{tbl:methods} presents and describes the forecasting methods (i.e., members) used for ensemble construction. Many of them are available on the NASA Community Coordinated Modelling Center Flare Scoreboard that is located at \url{https://ccmc.gsfc.nasa.gov/challenges/flare.php}. Four out of six methods (MAG4, ASSA, ASAP, MCSTAT) are fully automated, while the remaining two (NOAA, MOSWOC) are considered as human-influenced -- i.e., the raw forecasts, produced by a trained software, are adjusted according to a human forecaster's expertise and knowledge. All methods listed in Table~\ref{tbl:methods} produce forecasts at least every 24 hours, and forecast probabilities consist of the likelihood ($0-1$ being the decimal representation of $0-100$\%) of observing a flare of given class within the forecasting window, $\Delta t$.  A time span of three years (2014, 2015, and 2016) was considered in this study. This particular time period was chosen in order to maximize both the number of methods to be included and the number of forecasts without significant gaps in the data. 

\begin{table}[!t]
\begin{scriptsize}
\caption{Flare Forecasting methods included in the ensemble forecast (members). Name, developer/issuer/responsible institution, details on predictive model, archive or place used to retrieve forecasts, and references for each method are presented.}
\centering
\begin{tabular}{l l l l l}
\hline
Method       & Issuer/Responsible   & Predicting method     & Source & Reference \\
\hline
MAG4         & U. of Alabama,       & Forecasting curve +    & iSWA\footnotemark & \citet{Falconer2011,Falconer2014}\\
             & MSFC                 & Free energy proxy;  &  & \\
             &                      & fully automated     &  & \\
ASSA         & Korean Space         & McIntosh class +     &  iSWA & ASSA Manual\footnotemark\\
			 & Weather Center       & Poisson statistics; & & \\
             &                      & Fully automated. & & \\
ASAP         & U. of Bradford, UK   & McIntosh class;     & iSWA & \citet{Colak2008,Colak2009}\\
			 &                      & sunspot-group area; &  & \\
             &                      & Neural network.      &  & \\
NOAA         & NOAA SWPC            & Table look-up +      & \url{swpc.noaa.gov} & \citet{Crown2012}\\
             &                      & persistence +        &  & \\
             &                      & Climatology;        &  & \\
             &                      & human corrected.     &  & \\
MOSWOC       & UK Met Office        & McIntosh class +     & \url{metoffice.gov.uk/} & \citet{Murray2017} \\
             &                      & Poisson statistics; & \url{space-weather}  & \\
             &                      & human corrected     &    & \\
MCSTAT          & Trinity College      & McIntosh class +     & \url{solarmonitor.org} & \citet{Gallagher2002}, \\
             & Dublin               & Poisson statistics;  &    &  \citet{Bloomfield2012} \\
             &                      & fully automated.     &    & \\
\hline

\end{tabular}

{$^{1}$\url{iswa.ccmc.gsfc.nasa.gov}}\newline
{$^{2}$\url{http://spaceweather.rra.go.kr/images/assa/ASSA_GUI_MANUAL_112.pdf}}
\label{tbl:methods}
\end{scriptsize}
\end{table}

It is important to highlight that in order to combine the forecasts from different sources, these need to correspond to the same forecast window duration. For all methods but one, forecasts correspond to a 24-hour window. For the exception, ASSA, $\Delta t$ = 12 hours. In this case, because of the Poisson-statistics nature of that method, a 12-hour forecast can be transformed into a 24-hour forecast as illustrated in \citet{Guerra2015}. In addition, for methods such as MCSTAT and ASAP, which provide forecasts for individual active regions, the full-disk forecasts can be calculated according to \citet{Murray2017},

\begin{equation}
P_{\rm fd} = 1-\Pi_{i}(1-P_{i}) \ ,
\end{equation}

where $P_{i}$ is the probability of flaring for the $i$-th active region present on the disk. The product is taken over the total number of regions at the forecast issue time.

\begin{figure}[t!]
\centering
\includegraphics[width=13pc]{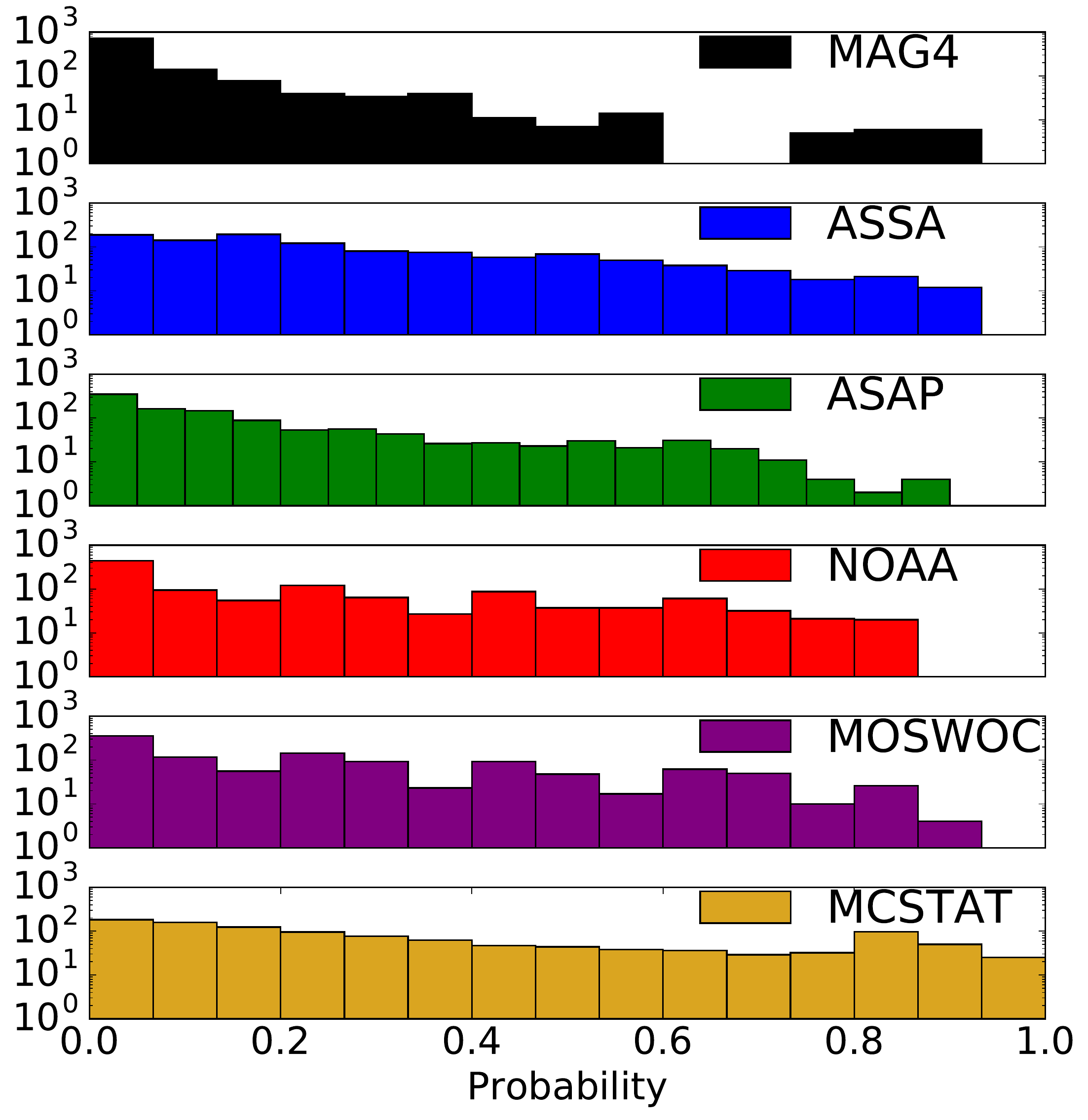}
\includegraphics[width=24pc]{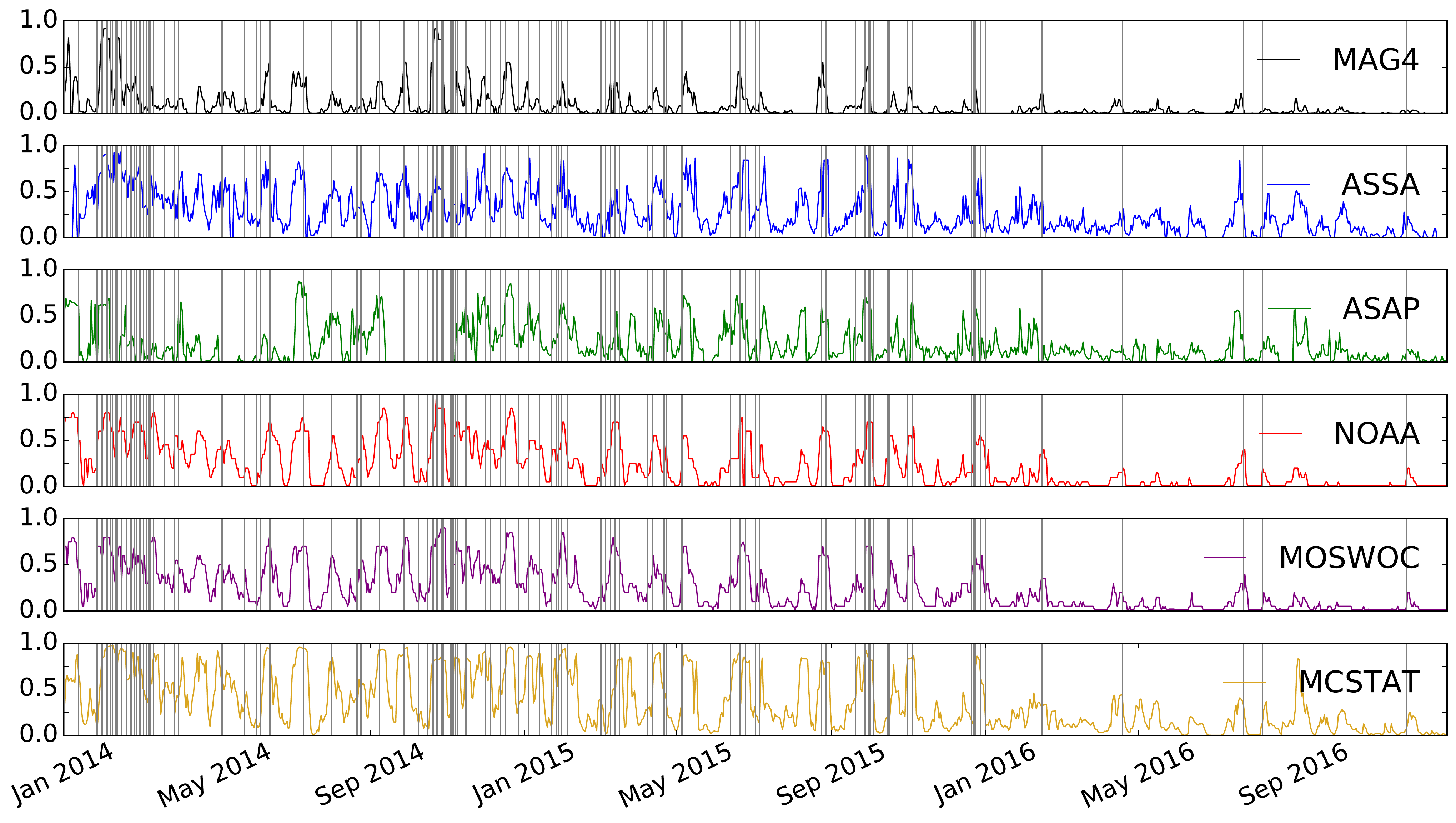}
\caption{Data sample. Probabilistic forecasts and events for M-class flares (histograms, left panels; time series, right panels). From the top, forecasting methods (color) are: MAG4 (black), ASSA (blue), ASAP (green), NOAA (red), MOSWOC (purple), and MCSTAT (gold). In the right panels, vertical grey lines signal positive events, i.e., days when at least one M-class flare was observed.}
\label{fig:m_forecasts}
\end{figure}

Major flares of GOES M- and X-classes were studied here, since C-class flare forecasts are not typically issued by operational centres. Figure~\ref{fig:m_forecasts} presents the 24-hour probabilistic forecast data for M-class flares, including histograms of values (left panels) and the full 3-year time series (right panels). Data is color-coded according to the forecasting method -- from top to bottom, black corresponds to MAG4, blue to ASSA, green to ASAP, red to NOAA, purple to MOSWOC, and gold to MCSTAT. All forecasts for M-class flares show similar characteristics -- probability values range from almost $0.0$ to $0.9 - 1.0$, with a decreasing frequency from low to high probability bins. Although, in case of MAG4, higher frequency is concentrated in the lowest-probability bin while some bins are empty. On the other hand, forecasts for X-class flares (not displayed) show a variety of upper limit for probability ranges -- between $0.25$ (ASSA) and $0.90$ (ASAP). 

During the study time period (2014-2016) a total of 18 X-class flares and 348 M-class flares were observed. However, due to the definition of forecasts stated above, the definition of events corresponds to the days in which at least one flare, of a particular class, was observed. Therefore, a time series of events is constructed by assigning 1.0 (positive) to flaring days and 0.0 (negative) otherwise. Since multiple flares can occur during the same day, the number of event days is not equal to the number of flares observed. Event days are displayed in the right-hand panels of Figure~\ref{fig:m_forecasts} by vertical gray lines. In total, 189 and 17 days between 1 January 2014 and 31 December 2016 have M- and X-class flares, respectively, yielding climatological event-day frequencies of 0.172 and 0.016, respectively.

\begin{table}[!t]
\caption{Matrix of Pearson's R correlation coefficients calculated among M-class flare forecast time series shown in Figure~\ref{fig:m_forecasts}, right panel. The mean value for each method is calculated using all the non-zero values in the column and row that corresponds to such method.}
\centering
\begin{tabular}{c|ccccc|c}
\hline
 &  ASSA & ASAP & NOAA & MOSWOC & MCSTAT & Mean \\
\hline
MAG4 & 0.615 & 0.431 & 0.689 & 0.718 & 0.653 & 0.621 \\
ASSA & -- & 0.534 & 0.661 & 0.705 & 0.769 & 0.657 \\
ASAP & -- & -- & 0.476 & 0.512 & 0.543 & 0.499 \\
NOAA & -- & -- & -- & 0.938 & 0.835 & 0.720 \\
MOSWOC & -- & -- & -- & -- & 0.849 & 0.744 \\
MCSTAT & -- & -- & -- & -- & -- & 0.730 \\
\hline
\end{tabular}
\label{tbl:corr_vals}
\end{table}

Visual inspection of the time series in Figure~\ref{fig:m_forecasts} reveals a certain level of correlation across all forecasting methods. This observation is not unexpected since all methods use parameterization of the same photospheric magnetic field or sunspot-related data as a starting point. Table~\ref{tbl:corr_vals} displays the linear correlation (Pearson's R) coefficients calculated between pairs of forecasting methods using the time-series data in Figure~\ref{fig:m_forecasts} (right panels). The last column on Table~\ref{tbl:corr_vals} shows the average correlation value across all methods -- average of all non-zero entries in the table for the column and row corresponding to the same method. Average correlation coefficients for M-class flare forecasts range between $\sim$0.50 and 0.75.  For X-class flare forecasts, average correlation coefficients are between $\sim$0.42 and 0.60. Although strongly-correlated methods could be considered as `repeated' forecasting information, it will be shown in the results of this investigation that their contributions to ensembles will basically depend on the overall forecasting accuracy.

\section{Ensemble Models}
\label{Section:Models}

Given a group of $M$ probabilistic forecast time series, all corresponding to the same type of event, a combined or {\it ensemble} prediction can be obtained by linear combination (LC) as \citet{Guerra2015},

\begin{equation}
P^{c}(\{w_{i}\}, \{P_{i}\}; t) = \sum_{i=0}^{M-1}w_{i}P_{i}(t) \ ,
\label{eq:lc_const}
\end{equation}

in which the index $i$ corresponds to the $i-th$ member in the group $\left\{P_{i}; i=0, \ldots, M-1\right\}$. The combining weight, $w_{i}$, determines the contribution of each member time series (i.e., forecasting method) to the ensemble prediction. This problem is reduced to finding an appropriate set of combination weights $\left\{w_{i}; i=0, \ldots, M-1\right\}$ that makes the ensemble prediction more accurate than any of the individual ensemble members. Three particular options to determine the combination weights are explored in this investigation: 1) error-variance minimization (performance history); 2) constrained metric optimization; 3) unconstrained metric optimization. Each of these options are explained in the following sections.


\subsection{Performance History}{\label{sec:performance_h}}

The simplest and most straightforward way to determine the set of combination weights is by looking at the performance history of each member \citep{Armstrong2001}. By doing this, higher weights are assigned to members with relatively good forecasting track record and lower weights to forecasts with poor performance \citep{Genre2013}.  Given that each forecast time series consist of the same number and range of discrete times, weights can be calculated as \citep{Stock2004},

\begin{equation}
w_{i} = \frac{m_{i}^{-1}}{\sum_{j=0}^{M-1}m_{j}^{-1}} \ ,
\label{eq:t_weights}
\end{equation}

where each member's weight is proportional to the reciprocal of their $m_{i}$, the cumulative sum of past partial errors,

\begin{equation}
m_{i} = \sum_{k=0}^{N-1}(P_{i,k}-E_{k})^{2} \ .
\label{eq:mi}
\end{equation}

In the Equation above, $E_{k}$ is the events time series and the index $k$ labels the discrete time range, $\{t_{k}; k = 0, \ldots,  N-1\}$. Equation~\ref{eq:mi} corresponds the unnormalized Brier Score -- the Mean Squared Error (MSE) for probabilistic forecasts -- since the $(N-1)^{-1}$ normalization coefficient cancels out because of the ratio in Equation~\ref{eq:t_weights}. Equation~\ref{eq:mi} implies that members with smaller partial error have larger weights. On the other hand, from Equations~\ref{eq:t_weights} and \ref{eq:mi} is easy to prove that,

\begin{equation}
\sum_{i=0}^{M-1}w_{i} = 1 \ ,
\label{eq:const}
\end{equation}

with $w_{i}>$0. This means that combination weights are {\it constrained} to add up to unity. This is important when the forecasts are probabilistic -- the value of $P^{c}$ cannot exceed 1. 

It can also be seen from Equations \ref{eq:t_weights} and \ref{eq:mi} that combination weights depend on the timeseries (forecasts and events) temporal range and resolution.


\subsection{Metric-optimized Constrained Linear Combination}

Alternatively, an optimal set of combination weights can be found by solving the optimization problem, 

\begin{equation}
\frac{d}{dw_{i}}\mathbb{M}(P^{c},E) = 0 \ , \quad i=0, \ldots, M -1 \ ,
\label{eq:opt_eqs}
\end{equation}

where $\mathbb{M}$ corresponds to a performance metric (a sort of loss function that quantifies the difference between forecasts and events), and $P^{c}$ is the linear combination given by Equation~\ref{eq:lc_const}. In this case the solution to Equation~\ref{eq:opt_eqs}, $\left\{w_{i}^{\rm{con}}\right\}^{\star}$, must also satisfy the constraint given in Equation~\ref{eq:const}. When using combination weights as described in this section and Section \ref{sec:performance_h}, the linear combination in Equation~\ref{eq:lc_const} is known as a {\it constrained linear combination} (CLC; \citet{Granger1984}).


\subsection{Metric-optimized Unconstrained Linear Combination}

On the other hand, an {\it unconstrained linear combination} (ULC; \citet{Granger1984}) of ensemble members' weights, $w^{\mathrm{unc}}_{i}$, can be constructed by adding a weighted contribution of the climatological frequency in as an additional probabilistic forecast. This results in the linear combination of Equation~\ref{eq:lc_const} becoming,

\begin{equation}
P^{c}(\{w_{i}\},\{P_{i}\};t) = \sum_{i=0}^{M}w^{\mathrm{unc}}_{i}P_{i}(t) + w_{E}\bar{E}(t) \ ,
\label{eq:lc_unconst}
\end{equation}

where $\bar{E}(t)$ is a time series with a constant value equal to the climatological frequency (calculated over the studied time period), and $w_{E}$ is its combination weight. In this case, Equation~\ref{eq:const} becomes,

\begin{equation}
\sum_{i=0}^{M}w^{\mathrm{unc}}_{i} + w_{E} = 1 \ ,
\label{eq:unconst}
\end{equation}

with ${w_{i}}^{\rm{unc}}$ and $w_{E}$ capable of taking positive or negative values. In this case, the sum of the combination weights for the ensemble's members (i.e. without $w_{E}$) is not constrained to any value. Hence, this particular linear combination is called unconstrained. Solving Equation~\ref{eq:opt_eqs} with Equations~\ref{eq:lc_unconst} and \ref{eq:unconst} provides a different group of ensembles given the optimal set of unconstrained weights, $\left\{w^{\mathrm{unc}}_{i}\right\}^{\star}$. In this case, $\bar{E}(t)$ functions as a benchmark level that takes into account the level of flaring activity over the three year time period studied here.


\section{Results}
\label{Section:Results}

In order to solve the optimization problem of Equation~\ref{eq:opt_eqs} (using either constrained or unconstrained linear combinations) and thus find the combination weights, a metric or loss function must be used. The constructed ensemble forecasts will be different from each other as much as the metrics are intrinsically different. The list of metrics employed in this work are presented in Table~\ref{tbl:metrics}. Probabilistic metrics are used as well as the more traditional categorical metrics \citep{Murray2017}, although ensembles methods are versatile enough to fulfill the requirements of operational environments by allowing the use of any metric that might be of particular interest.

\begin{table}[!t]
\caption{Performance metrics tested in the optimization process. Each metric produces a different set of combination weights (i.e., a different ensemble). In each case a label is shown in parentheses that is used throughout the rest of the manuscript. Categorical metrics are calculated using  2$\times$2 contingency table after probabilistic forecasts are turned into deterministic forecasts by choosing a threshold value, $P_{\rm th}$. See Appendix A for their definitions.}
\centering
\begin{tabular}{l l}
\hline
Probabilistic                        & Categorical    \\
\hline
Brier score (BRIER)                  &  Brier score (BRIER\_C)\\
Mean absolute error (MAE)            &  True skill statistic (TSS)\\
Linear correlation coefficient (LCC)      &  Heidke skill score (HSS)\\
Rank (Nonlinear) correlation coefficient\footnotemark  &  Accuracy (ACC)\\
(NLCC\_$\rho$, NLCC\_$\tau$)         &  Critical success index (CSI)\\
Reliability (REL)                    &  Gilbert skill score (GSS)\\
Resolution (RES)                     &  \\
Relative Operating Characteristic (ROC) curve area               &  \\
\hline
\end{tabular}
\label{tbl:metrics}
\\
{\footnotesize
{$^{3}$NLCC\_$\rho$ and NLCC\_$\tau$ are Spearman's rank correlation and Kendall's $\tau$ correlation, correspondingly.}
}
\end{table}


\subsection{Combination Weights}\label{sec:comb_w}

Equation~\ref{eq:opt_eqs}, along with the corresponding constraint of Equation~\ref{eq:const} or \ref{eq:unconst}, was solved with the {\it Scipy} optimization software \citep{Oliphant2007} using a Sequential Least SQuares Programming \citep[SLSQP;][]{Kraft1988} solver method. SLSQP is an iterative method for solving nonlinear optimization problems with and without bounded value constraints. Initial-guess values for $\{w_{i}\}$ are provided to the routine, while the derivatives (with respect to the weights) are calculated numerically. The SLSQP method only performs a minimization of the function value, therefore for those metrics in Table~\ref{tbl:metrics} that require maximization (e.g., LCC, NLCC, ROC), the negative value of the metric is used as a function to minimize. For some of the optimization metrics, the resulting weights showed sensitivity to the initial-guess values given to the SLSQP solver -- possibly due to the metric being noisy at the resolution of the solver. Therefore, in order to ensure that the solution $\left\{w_{i}\right\}^{\star}$ corresponds to a global minimum, for each ensemble the solver is executed 500 times with randomly-selected initial values between, [0,1] for constrained case and [-1,1] for unconstrained case, at every step. This results in a distribution of values for each weight. In most cases, these distributions (not shown here) are normal in shape, therefore the mean value is used as the final optimized weight, with standard error (deviation) associated of up to $\sim$10\% of the mean value. However, in few cases, distributions appeared wider due to the noisy nature of the metric (loss) function.

In the following sections, only the results for M-class flare events are presented and discussed with mention to the results for X-class flares. Corresponding plots for X-class flare events can be found in Appendix B. It is worth keeping in mind that, due to the relatively low number of X-class event days, results for M+ (i.e., flares of M-class and above) will be similar to those for only M-class flares because these flares will dominate the statistics in the sample used.

\begin{figure}[!t]
\centering
\includegraphics[width=37pc]{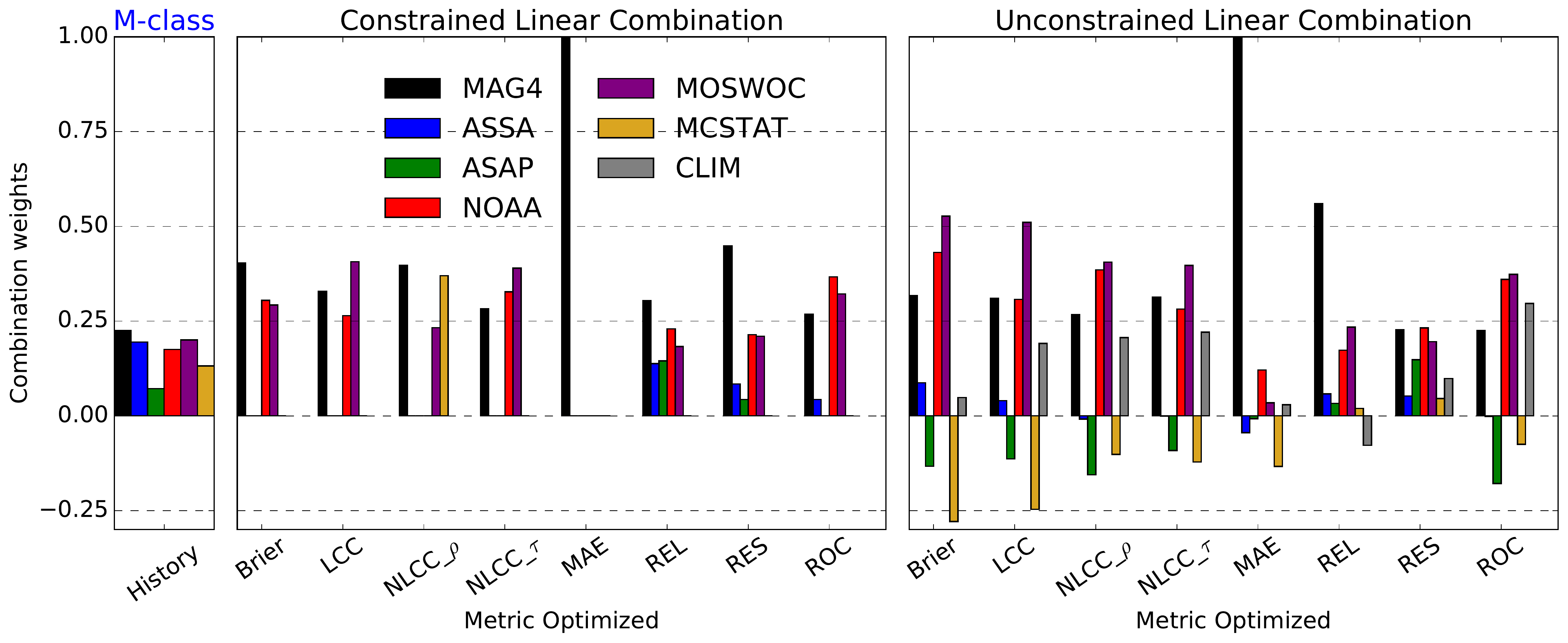}
\caption{Ensemble combination weights for the optimization of probabilistic metrics (Table~\ref{tbl:metrics}, left column) on M-class flare forecasts. {\it Left} panel corresponds to combination weights calculated from performance track (see text for details) while {\it Middle} and {\it Right} panels correspond to constrained and unconstrained linear combinations, respectively. Weights are presented using the same color scheme as Figure~\ref{fig:m_forecasts} for each forecast method member. Note, ULC weights corresponding to the climatological forecast member are displayed in gray in the right panel.}
\label{fig:w_prob_m}
\end{figure}

Figure~\ref{fig:w_prob_m} shows the optimized combination weights $\{w_{i}\}^{\star}$ for the performance history (left panel) and probabilistic metrics (outlined in Table~\ref{tbl:metrics}). Middle and right panels correspond to the constrained ($\{w_{i}^{\mathrm{con}}\}^{\star}$) and unconstrained ($\{w_{i}^{\mathrm{unc}}\}^{\star}$) linear combinations. Combination weights are displayed according to the color code used in Figure~\ref{fig:m_forecasts}. It is shown in the right panel that for the ULC case some combination weights acquire negative values, as expected. It is worth noting that negative values do not necessarily imply that such a member/method performance is worse than those members with positive weights because it is this particular combination that is necessary to optimize the chosen metric. It is clear that ensembles (i.e., the sets of combination weights) are generally very different for the optimization of differing metrics and the type of linear combination. However, some general characteristics are observed: 1) human-adjusted members appear in most ensembles with major (positive) contributions -- i.e., larger magnitudes than the equal weighting values of $w^{\mathrm{con}}_{\mathrm{eq}} = 1/6 = 0.167$ and $w^{\mathrm{unc}}_{\mathrm{eq}} = 1/7 = 0.142$ for the CLC and ULC cases, respectively; 2) combination weights for members that are zero in the CLC case tend to show negative values in the ULC case; 3) for most ULC ensembles, the climatological forecast member has a positive weight, implying that for the ensemble members considered and the time range studied, the level of activity might have been underforecast by some of the members.

It is also clear from Figure~\ref{fig:w_prob_m} that using the ULC approach results in the formation of ensembles with more members having non-zero weights (i.e., more diverse ensemble membership). For X-class flare events, the resulting ensembles are more sensitive to the metric used (see Fig.~\ref{fig:w_prob_x}). No clear tendency arises in that case, however these results seem highly dependent on the low number of X-class flares in the studied sample.

\begin{figure}[!t]
\centering

\includegraphics[width=34pc]{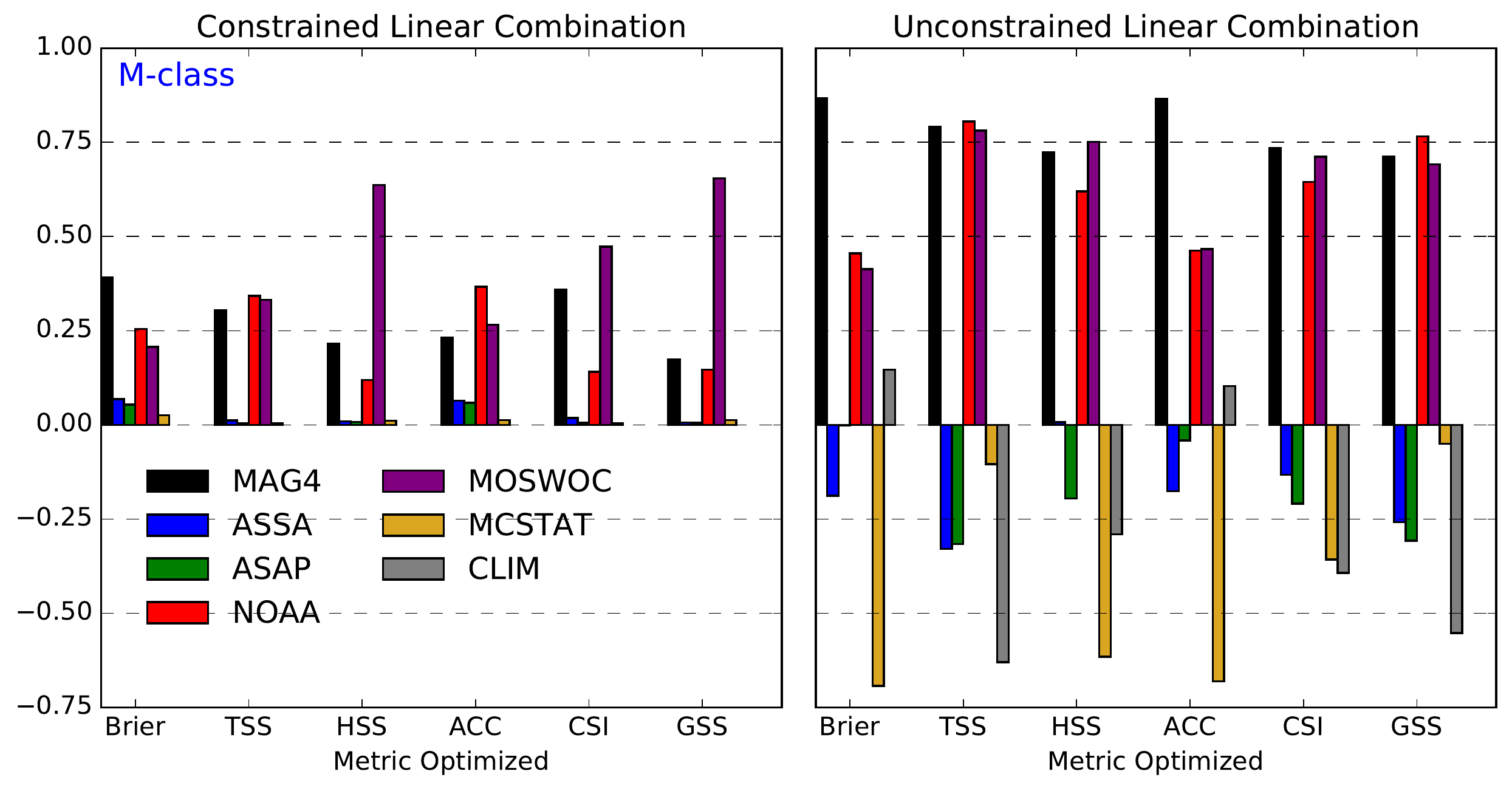}
\caption{Similar to Figure~\ref{fig:w_prob_m}, but for the optimization of categorical metrics (Table~\ref{tbl:metrics}, right column).}
\label{fig:w_cat_m}
\end{figure}

Figure~\ref{fig:w_cat_m}, on the other hand, corresponds to the categorical-metric counterpart of Figure~\ref{fig:w_prob_m}. For categorical metrics, the threshold value used to transform probabilistic forecasts to deterministic forecasts is determined during the optimization process. See Appendix A for details about this thresholding procedure. For categorical-metric ensembles, it is observed: 1) unlike probabilistic-metric ensembles, weights determined using the CLC approach seem to consistently show non-zero values for most metrics; 2) for both CLC and ULC, ensembles seem more similar to each other in terms of the combination weights (i.e., the same members appear to dominate in most ensembles, being MAG4, NOAA, and MOSWOC); 3) weights for the climatological forecast member appear with negative values in most ensembles, contrary to the probabilistic case.


\subsection{Optimized Metrics}\label{sec:opt_metrics}

\begin{figure}[!t]
\centering

\includegraphics[width=18pc]{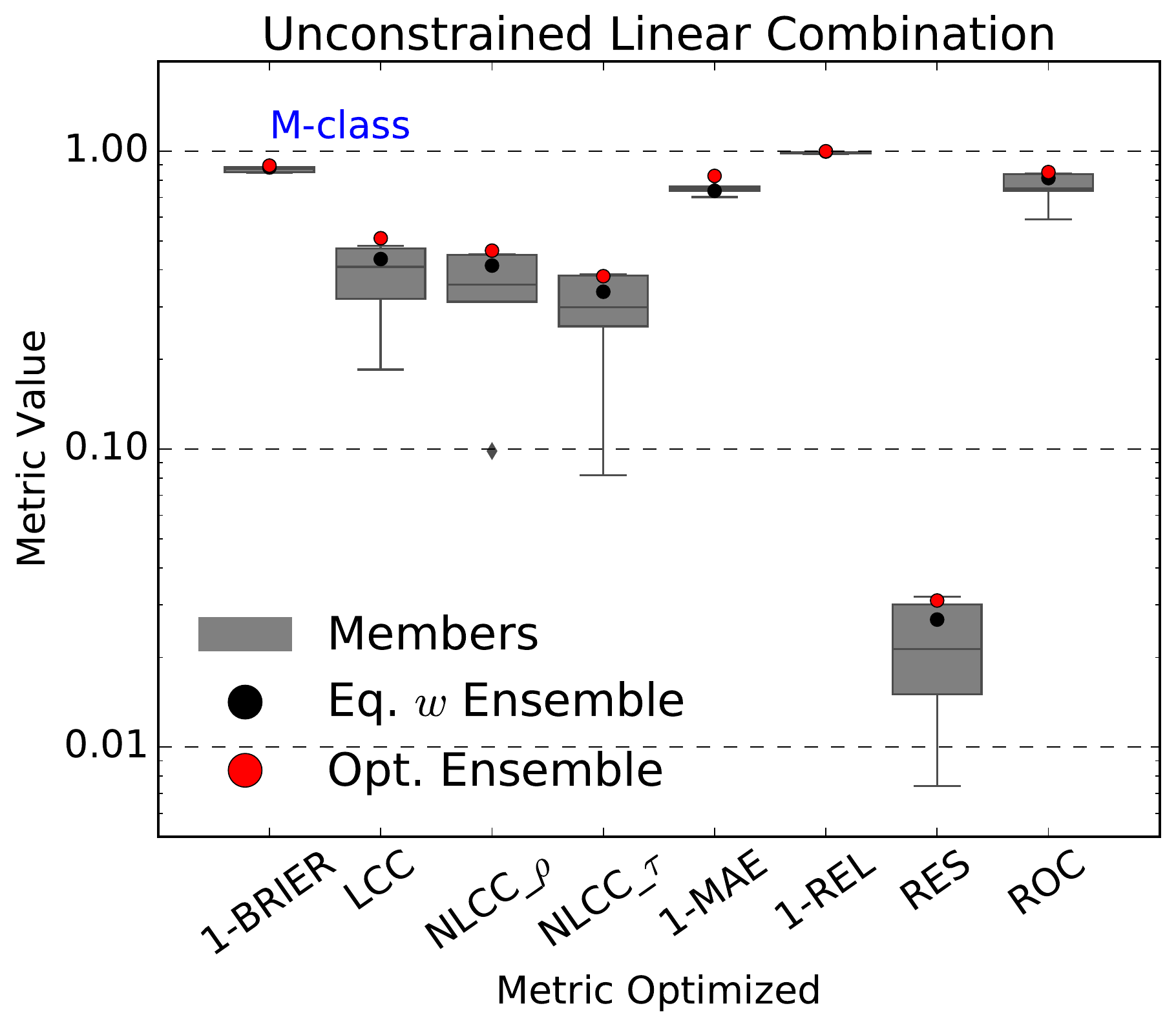}
\includegraphics[width=18.5pc]{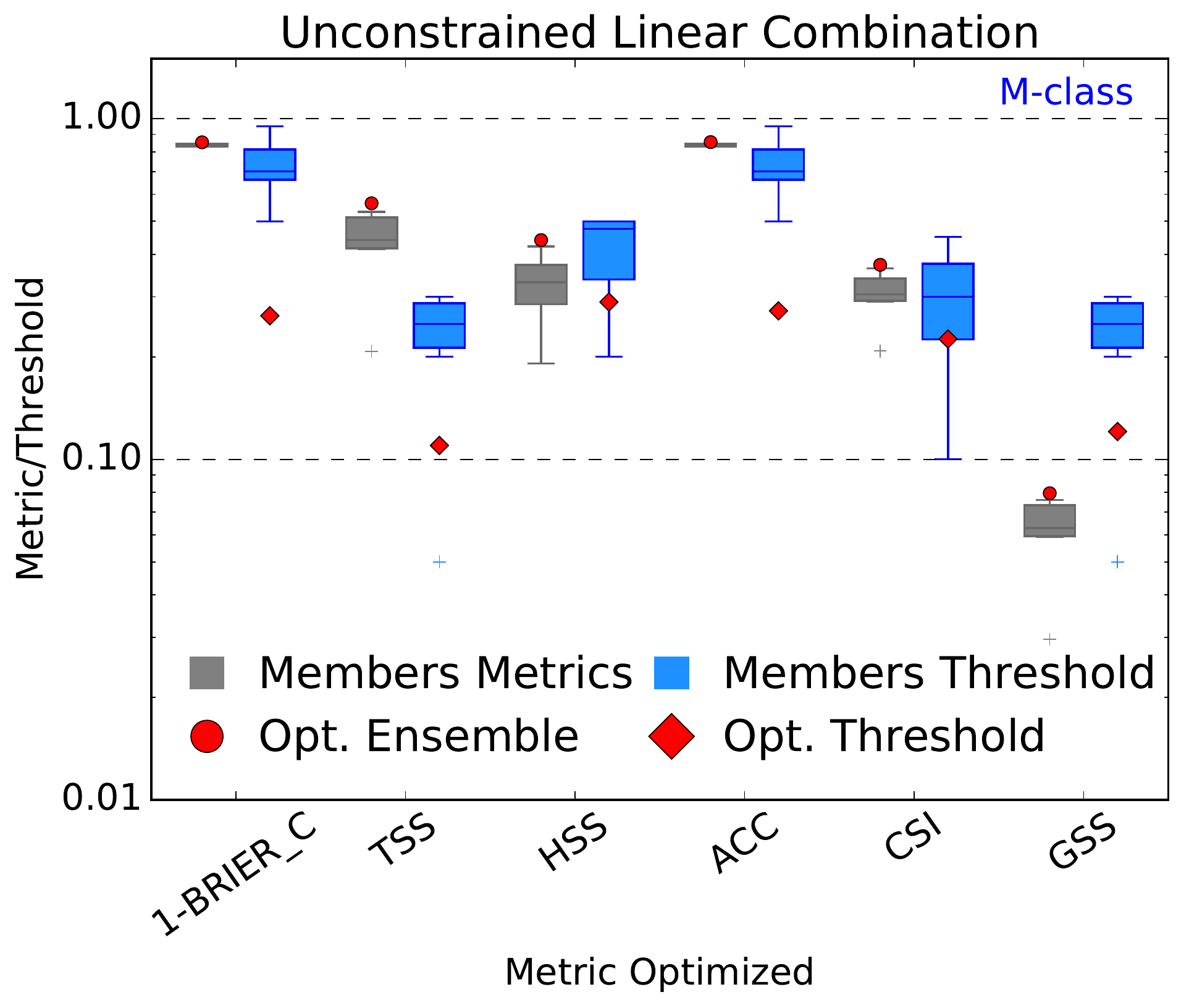}
\caption{{\it Left:} For probabilistic metrics three values are shown per metric: 1) metrics values for ensemble members are displayed as box-and-whiskers; 2) the metric of the equal-weights ensemble (black circle); 3) the optimized (or best performing) ensemble (red circle). {\it Right:} Metric and threshold values for categorical-metric ensembles. Gray and blue box-and-whiskers plots correspond to the ensemble members' metrics and thresholds. Red circles and diamonds correspond to the optimized-ensemble metrics and thresholds. For better comparison, metrics in both panels that require minimization (i.e., BRIER, MAE, and REL) are displayed as $1 - $~(metric value).}
\label{fig:metrics_m_unconst}
\end{figure}

As indicated above, in order to determine combination weights such as those in Figures~\ref{fig:w_prob_m} and \ref{fig:w_cat_m}, the value of a chosen metric is optimized. In Figure~\ref{fig:metrics_m_unconst} these optimized metric values are presented for M-class flare forecasts using the ULC approach. The left panel corresponds to probabilistic-metric-optimized ensembles, while the right panel shows the categorical-metric-optimized ensembles. 

For the probabilistic metrics (Fig.~\ref{fig:metrics_m_unconst}, left panel), several values are presented: grey box-and-whiskers show the individually-calculated metrics for all members (top and bottom of the box represent first and third quartiles, the horizontal line in between correspond to the median, and the whiskers signal maximum and minimum); a metric value for the equal-weights ensemble (arithmetic mean; black circle), and the value for the best-performing ensemble (red circle; using the weights from Figure~\ref{fig:w_prob_m}, {\it Middle}). For a more convenient visualization, those metrics that are minimized (i.e., BRIER, MAE, and REL), are displayed as $1 - $~(metric value). In this way, better performing metric values are concentrated towards the upper limit (i.e., 1) of the range. For all the metrics in the left panel of Figure~\ref{fig:metrics_m_unconst}, the observed tendency is that the best-performing ensemble yields a metric value greater than that of the equal-weights ensemble ($\mathbb{M}$(Best-Perf. Ensemble) $>$ $\mathbb{M}$(Eq.-$w$ Ensemble)) which, in turn, produces a metric value greater than the median of the members' individual metric values ($\mathbb{M}$(Eq.-$w$ Ensemble) $>$ $\bar{\mathbb{M}}_{i}$). However, the equal-weights ensemble metric value often lies above the median value, meaning that one or two members perform better in this metric than the equal-weights ensemble. In addition, the best performing ensemble (red circle) often produces a metric value higher than the upper limit of the box and sometimes even higher than the maximum. This implies that the best-performing ensemble also often performs better than the best-performing individual member.

The right panel of Figure~\ref{fig:metrics_m_unconst} displays metric values and probability threshold values for the categorical metrics. In a similar way to the probabilistic-metrics case, box-and-whiskers correspond to the ensemble members' values, while symbols correspond to the values of the best-performing ensemble. Grey and blue box-and-whiskers correspond to metric and threshold values. The equal-weights ensemble is not shown here for practical reasons, but the results are similar to the probabilistic case. This plot shows two clear tendencies: 1) All categorical-metric-optimized ensembles achieved a metric value larger than the median for their members. Indeed, all red circles are seen outside the whiskers' range -- that is, the optimized-ensemble metric is larger than that of the best performing individual member; 2) the probability threshold values that are found to maximize the metrics are lower than the average probability threshold across the members.

In the case of X-class flares (Appendix Fig.~\ref{fig:w_cat_x}), results show similar tendencies to those described above. Also, for both the M- and X-class event levels, optimized metric values for the CLC case (not shown here) follow similar patterns as for the ULC case (i.e., $\mathbb{M}$(Opt. Ensemble) $>$ $\mathbb{M}$(Eq.-$w$ Ensemble) $>$ $\bar{\mathbb{M}_{i}}$). However, metric values for ULC ensembles are typically up to 5\% (M-class) and 15\% (X-class) higher than those of CLC ensembles. These results demonstrate that using the ULC approach when constructing ensembles achieves more optimal values for both probabilistic and categorical metrics. The improvement of metric values appeared larger for the X-class event level perhaps suggesting that ensembles might be very useful for rare events. However, with the low number of events available for this class, the statistical significance of such suggestion can't be shown.


\subsection{Ensembles Comparison}

The performance of how each ensemble output may be evaluated is demonstrated here using a variety of probabilistic validation metrics. It is important to clarify that the results in this section do not correspond to those of a validation process because the metrics were calculated in-sample. These results are intended to demonstrate that the choice of optimization metric for constructing an ensemble is fundamentally important/influential if the best-performing forecasts are to be desired. It is worth noting that `best performing' can mean different things depending on the end-user (see Section~\ref{Section:Conclusions} for further discussion), and here a selection of commonly-used metrics are used to simply showcase the usefulness of this technique.

Following the operational flare forecasting validation measures used by \citet{Murray2017} and \citet{Leka2019}, ROC curves and reliability diagrams are displayed in Figure~\ref{fig:verify_m} for a selection of M-class forecast ensemble members and final optimized ensembles (see Appendix \ref{fig:verify_x} for the equivalent X-class case). Reliability diagrams are conditioned on the forecasts, indicating how close the forecast probabilities of an event correspond to the actual observed frequency of events. These are good companions to ROC curves that are instead conditioned on the observations. ROC curves present forecast discrimination, and the area under the curve provides a useful measure of the discriminatory ability of a forecast. The ROC area for all forecasts as well as Brier score is presented in Appendix Tables~\ref{tbl:validation_metrics_m} and \ref{tbl:validation_metrics_x} for M- and X- forecasts, respectively. Brier score measures the mean square probability error, and can be broken down into components of reliability, resolution, and uncertainty, which are also listed in these tables.

For each table the scores are grouped in order of original input forecasts, ensembles from probabilistic-metric-optimization, and categorical-metric-optimization. Results for both CLC and ULC approaches are included. In general, the M-class forecast results are better than X-class, although that is to be expected considering the small number events in the time period used (only 17 X-class event days compared to 189 M-class event days out of 1096 total days). Most values are to be expected, with overall good Brier score but poor resolution, and only a few resulting forecasts with a `poor' ROC score (in the range 0.5 - 0.7). It is interesting to see in Table~\ref{tbl:validation_metrics_m} that overall the equal-weighted ensemble outperforms MAG4, which is the best of the automated (without human input) M-class forecasts, but that the human-edited MOSWOC forecast is the best performing overall in the original members group.

Group rankings are also included in both Appendix Tables~\ref{tbl:validation_metrics_m} and \ref{tbl:validation_metrics_x}, calculated by first ranking the forecasts based on all four scores separately, and then taking an average of the rankings and re-ranking for each group. Although the broad study of \cite{Leka2019} found that no single forecasting method displayed high performance over many skill metrics, this ranking averaging is done here in order to observe if there are major differences between probabilistic and categorical metrics. The top performers for each group in M-class forecasts are MOSWOC, NLCC\_$\rho$\_unc, and CSI, while for X-class the best performing forecasts are MOSWOC, LCC\_unc, and CSI. It is worth noting that rankings may change quite significantly depending on the metrics used, therefore the raw forecast data is freely provided to the reader to compare the results using any metric of their own interest (see Acknowledgements). Table~\ref{tbl:validation_metric_ranks_m} summarizes the top five performers for each metric based on their rankings separately. In this table both BRIER\_unc and NLCC\_$\rho$\_unc ensembles appear in the top five of all four evaluation metrics, while the LLC\_unc metric appears in evaluation skill metrics. Therefore, these three ensembles will be often be used as a sample of ``overall" top performers in the following sections.

For comparison purposes, Figures~\ref{fig:verify_m} and \ref{fig:verify_x} display ROC and reliability plots for a selection of these top-performing methods based on the rankings in the three groups. The upper row compares the best original ensemble members of each different method type and outputs, namely MOSWOC (human-edited, black line), MAG4 (automated, turquoise line), the equal-weights ensemble (blue line), and one of the top performing probabilistic ensembles as per Table~\ref{tbl:validation_metric_ranks_m}, NLCC\_$\rho$\_unc (purple line). The other rows compare forecasts within ranking groups, for example the middle row shows best constrained vs unconstrained probabilistic weighted methods, and the lower row constrained vs unconstrained categorical weighted methods. For the ROC curves in the left column, the better performing methods should be tending towards the upper left corner of the plot. For the reliability diagrams in the right column, methods should preferably be in the gray-shaded zone of `positive skill' around the diagonal, and if they are tending toward the horizontal line they are becoming comparable to climatology. 

These figures provide an easier illustrative depiction of the scores presented by the tables. For example, the ROC curve in the upper row of Figure~\ref{fig:verify_m} highlights the clear improvement that the ensembles have over the automated MAG4 method, with all other curves similarly good for M-class forecasts. The reliability diagrams of Figure~\ref{fig:verify_m} show that most methods/ensembles generally over-forecast (i.e., data points lie to the right of and below the center diagonal line), except the NLCC\_$\rho$\_unc ensemble. The plots for X-class forecasts in Appendix Figure~\ref{fig:verify_x} clearly highlight the issues related to rare-event forecasting, with poorer results across the board for all methods/ensembles compared to the M-class forecast results.

\begin{table}[!t]
\caption{Rankings of evaluation metrics for M-class flare forecasts. For each metric top five ensembles are displayed.}
\centering
\begin{tabular}{l c c c c}
\hline \hline
Rank 	& Brier 				& Reliability 		& Resolution 		& ROC 				\\
		& Score					&  					& 					& Area				\\
\hline \hline
1		& LCC\_unc				& NLCC\_$\rho$\_unc	& LCC\_unc			& NLCC\_$\tau$\_unc \\
2		& BRIER\_unc  			& BRIER\_unc    	& CSI\_unc			& NLCC\_$\rho$\_unc	\\
3		& NLCC\_$\rho$\_unc   		& BRIER\_C 		& HSS\_unc			& ROC\_unc			\\
4		& HSS\_unc				& GSS 				& BRIER\_unc		& BRIER\_unc		\\
5		& ROC\_unc				& REL\_unc 			& NLCC\_$\rho$\_unc	& LCC\_unc			\\	
\hline
\end{tabular}
\label{tbl:validation_metric_ranks_m}
\end{table}

\begin{figure}[!t]
\centering
\includegraphics[width=\textwidth]{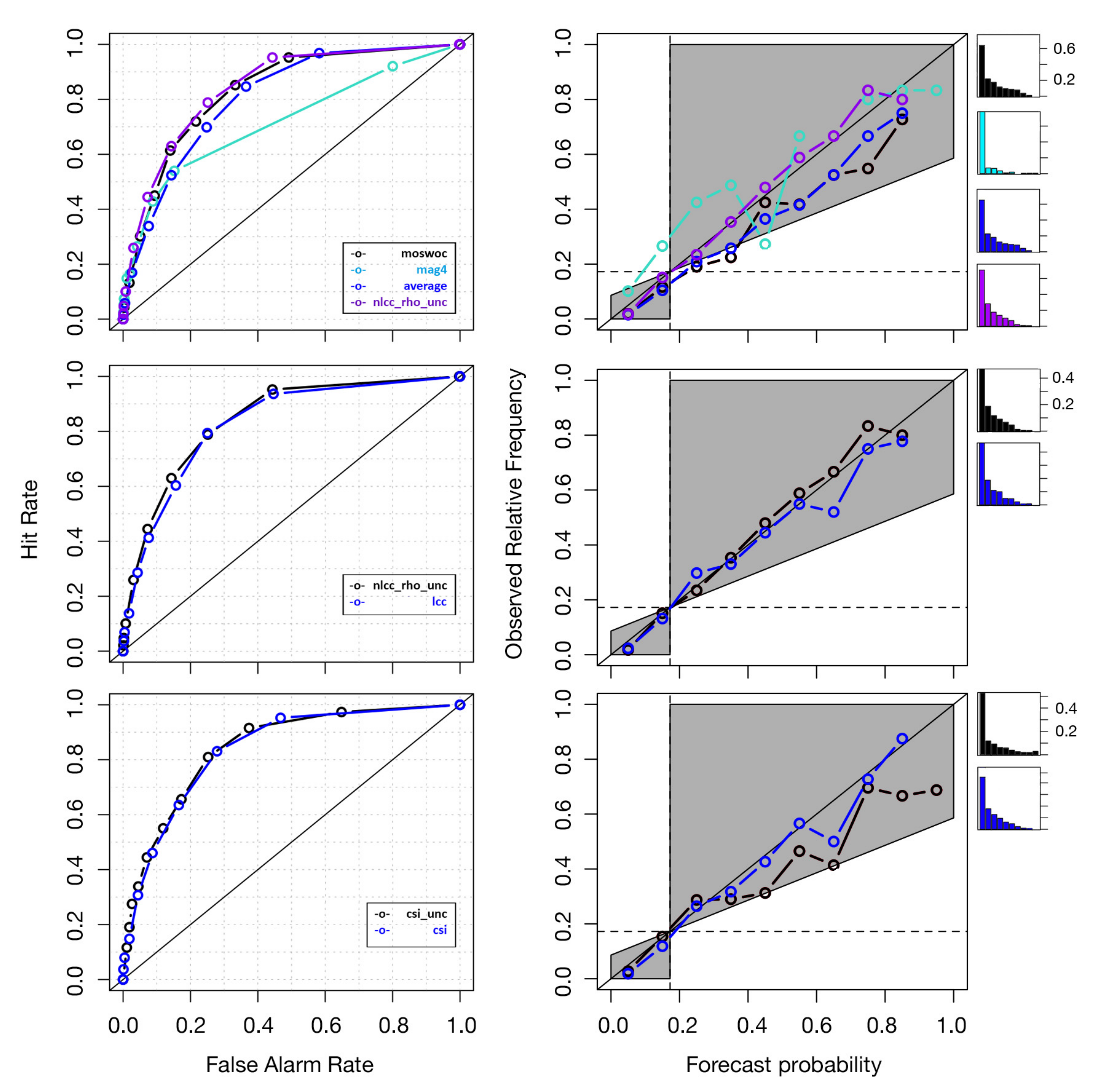}
\caption{ROC curves ({\it Left} column) and reliability diagrams ({\it Right} column) for M-class flare forecasts, comparing the top ranking individual method types and final ensemble performer ({\it Upper} row), and for constrained and unconstrained ensembles based on probabilistic ({\it Middle} row) and categorical ({\it Lower} row) metrics. Note that the center diagonal line in the ROC curves represents no skill, while for the reliability diagrams it indicates perfect reliability. The shaded areas in the reliability diagrams indicate regions that contribute positively to the Brier skill score (not shown/used here).}
\label{fig:verify_m}
\end{figure}


\subsection{ Ensembles Uncertainties}

There are two main uncertainty sources associated with the linearly-combined (ensemble) probabilistic forecasts $P^{c}$ here constructed: 1) the uncertainty associated with the weighted average; 2) systematic uncertainties associated with the input data for Equations~\ref{eq:lc_const} and \ref{eq:lc_unconst} (i.e., forecasts and weights). Thus,

\begin{equation}
u^{2}(P^{c}) = u_{\rm{stat}}^{2} + u_{\rm{syst}}^{2} \ ,
\label{eq:total_error}
\end{equation}

\noindent
where $u_{\mathrm{stat}}^{2}$ can be calculated as a weighted standard deviation. A simplified version (due to the constraints in Equations~\ref{eq:const} and \ref{eq:unconst}) of the standard error of the weighted mean (SEM) formulation presented by \citet{Gatz1995}

\begin{equation}
u_{\rm{stat}}^{2}  = \frac{M}{M-1}\sum_{i=0}^{M-1}w^{2}_{i}(P_{i}-P^{c})^{2} \ .
\label{eq:statistical}
\end{equation}

is adopted here. Equation~\ref{eq:statistical} corresponds to the typical SEM corrected by the factor $M/(M-1)$. On the other hand, error propagation through the linear combination  (Equations~\ref{eq:lc_const} and \ref{eq:lc_unconst}) will provide estimates for the uncertainties associated with the input data. In its most general form this is,

\begin{equation}
u^{2}_{\rm{syst}} = \sum_{i=0}^{M-1}\{w^{2}_{i}u^{2}(P_{i}) + P^{2}_{i}u^{2}(w_{i})\} \ ,
\label{eq:systematic_propagation}
\end{equation}

where $u(P_{i})$ in the first term is the uncertainty associated with the probabilistic forecasts for the i-th ensemble member and $u(w_{i})$ in the second term is the uncertainty of the i-th member combination weight. Most ensemble members in this study do not have uncertainties associated to their forecasts. As it was mentioned in Section \ref{sec:comb_w}, combination weights such as those in Figures~\ref{fig:w_prob_m} and \ref{fig:w_cat_m} (as well as Appendix Figs.~\ref{fig:w_prob_x} and \ref{fig:w_cat_x} for the X-class case) correspond to the mean values of normal distributions. Therefore their uncertainties can be represented by the corresponding standard deviation, $\sigma(w_{i})$. Since Equation~\ref{eq:systematic_propagation} implies that the more members the ensemble has the larger the uncertainty, the systematic errors must be normalized by the number of members with non-zero weights, $M'$. Therefore, Equation~\ref{eq:systematic_propagation} reduces to,

\begin{equation}
u^{2}_{\rm{syst}} = \frac{1}{M'}\sum_{i=0}^{M-1}P^{2}_{i}\sigma^{2}(w_{i}) \ .
\label{eq:u_systematic}
\end{equation}

\begin{figure}[!t]
\centering
\includegraphics[width=0.475\textwidth]{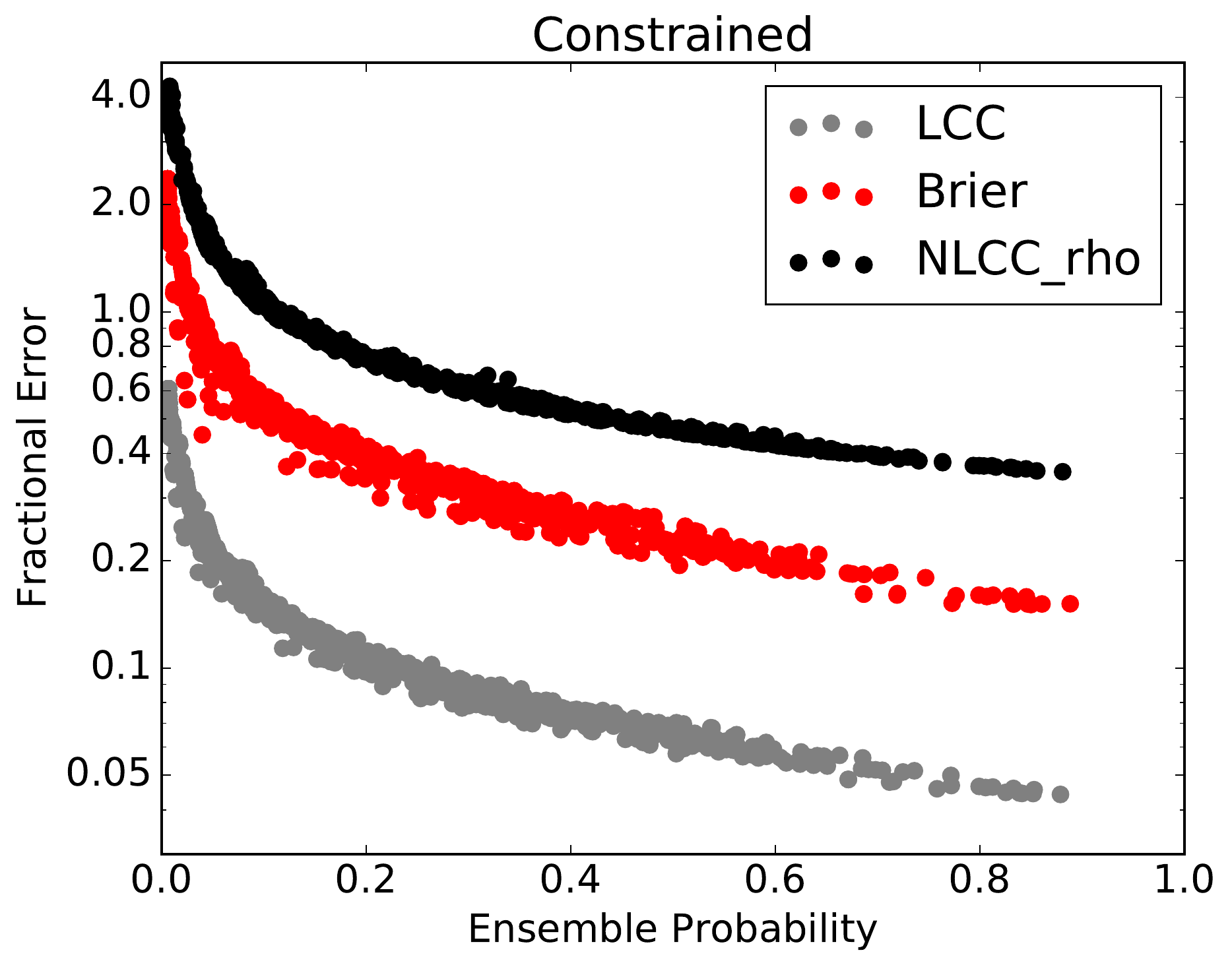}
\includegraphics[width=0.475\textwidth]{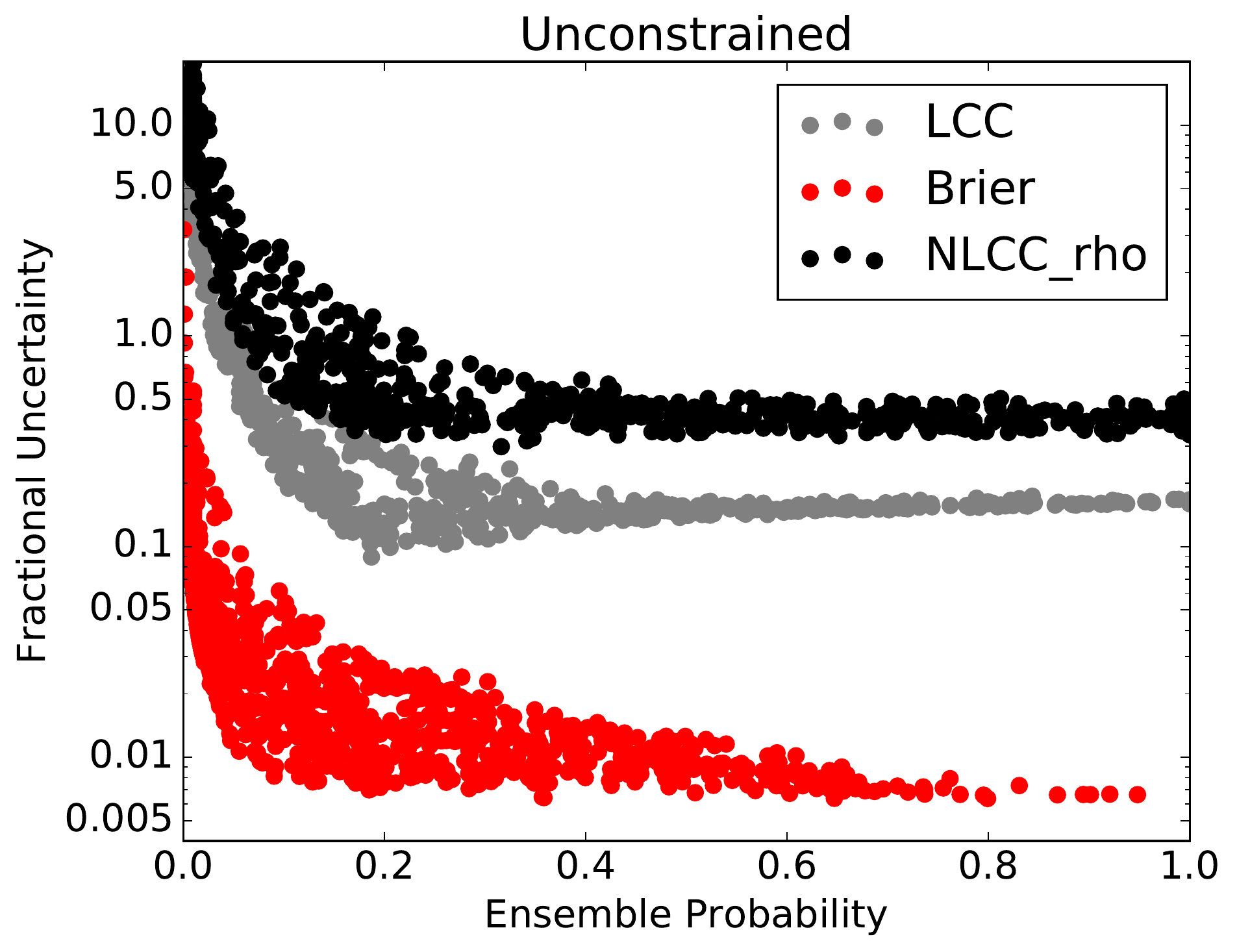}
\caption{Fractional uncertainties as a function of ensemble probability. {\it Left} and {\it Right} panels compare the CLC and ULC cases for the three top-performing ensembles (Table~\ref{tbl:validation_metric_ranks_m} for M-class flares, consisting of the metrics linear correlation coefficient (LCC; grey), Brier score (Brier; red), and non-linear rank correlation coefficient, $\rho$ (NLCC\_$\rho$; black).}
\label{fig:errors_M}
\end{figure}

Figure~\ref{fig:errors_M} displays the fractional errors calculated with Equations~\ref{eq:total_error}, \ref{eq:statistical}, and \ref{eq:u_systematic} for those three metrics that repeatedly appeared in Table~\ref{tbl:validation_metric_ranks_m}: LCC (grey), BRIER (red), NLCC\_$\rho$ (black). Left and right plots in Figure~\ref{fig:errors_M} compare the constrained (CLC) and unconstrained (ULC) cases for all three metrics. Uncertainties in both linear-combination cases (CLC and ULC) show a similar trend -- fractional errors are larger for low probabilities than for large probabilities. In the case of ULC (Fig.~\ref{fig:errors_M}, left panel), fractional errors appear to always decrease with increasing probability value in a slow, non-linear way. In this case, the LCC-optimized ensemble provides the lowest errors with values ranging from 5\% when $P \rightarrow 1.0$ to 60\% when $P \rightarrow 0.0$. On the other hand, ULC ensembles show a slow non-linear decrease at low probability values, but then fractional errors seem to reach a constant levels. In this case, the BRIER-optimized ensemble gives the overall lowest errors ranging between 0.5\% and 5\% when $P \gtrsim$ 0.2.

There are two important aspects to consider regarding the uncertainties for the ensemble models presented here. First, the uncertainty term associated with the combination weights (Equation~\ref{eq:u_systematic}) can also consider the uncertainty of each individual set of weights that makes up the distribution, that is $u^{2}(w_{i}) = \sigma^{2}(w_{i}) + u^{2}(\overline{w}_{i})$. This contribution can be calculated by error propagation through the mean-value expression. However, the uncertainty $u^{2}(\overline{w}_{i})$ term is not included in the present results since the SLSQP solver does not provide the such uncertainties directly. Second, the uncertainty values presented here (i.e. Figure \ref{fig:errors_M}) were calculated with both weights and forecasts in-sample. For out-of-sample uncertainties, a similar behavior can be expected. The total uncertainty $u(P^{c})$ grows with increasing probability $P_{i}$ value making the fractional uncertainty decrease as seen in Figure \ref{fig:errors_M}. The values of weights (Equation \ref{eq:statistical}) and their uncertainties (Equation \ref{eq:u_systematic}), which are always calculated in-sample, should not affect this specific behavior, instead they should only determine the rate of growth and overall level of uncertainties.

\section{Concluding Remarks}
\label{Section:Conclusions}

This investigation presented the modeling and implementation of multi-model input ensemble forecasts for major solar flares. Using probabilistic forecasts from six different forecasting methods that are publicly available online in at least a `near-operational' state, three different schemes for linearly combining forecasts were tested: track history (i.e., variance minimization), metric-optimized constrained, and metric-optimized unconstrained linear combinations. In the last two cases, a group of 13 forecast validation metrics (7 probabilistic and 5 categorical) were used as functions to be optimized and thus find the most optimal ensemble combination weights. Resulting ensemble forecasts for this study time (2014--2016, inclusive) were compared to each other and ranked by using four widely used probabilistic performance metrics: Brier score, reliability, resolution, and ROC area. Finally, uncertainties on each ensemble were studied.

A total of 28 ensembles were constructed to study M- and X-class flare forecasts. The vast majority of ensembles not only performed better -- as measured by the four metrics -- than all the ensemble members but also better than the equal-weights ensemble. This means that even though a simple geometric average of forecasts will be a more accurate forecast than any one of the original ensemble members on their own, according to the results in this investigation, that is not necessarily the most optimal linear combination. For both flare event levels, different optimization metrics lead to differing ensemble combination weights with non-zero weights from both automated and human-influenced members. When the combination weights are forced to have only positive values (i.e., a constrained linear combination) it is observed that optimization of the more mathematical metrics (i.e., Brier, LCC, NLCC, and MAE) do not necessarily include all members, in contrast to optimization of the more attribute-related metrics (i.e., RES, REL, ROC). When the ensemble member weights are allowed to obtain negative values as well, those previously zero-weighted members are observed to have negative weights. It is important to highlight that a negative weight does not mean that it is less important than a positive-valued weight since it is the overall linear combination (positive and negative weights) that achieves the optimal metric value. The tendency is similar for all categorical metrics.

The optimized combination weights provided final metric values greater than both the metric calculated using an equal-weights combination and the average metric value across all members. However, only in the M-class case with a greater number of event days in the time period studied did every optimized ensemble show a metric value better than all of the ensemble members. As it is expected, the relatively low number of X-class event days in our data sample is not enough to make every optimized ensemble better than any of the members. It is in these cases where the choice of optimizing metric is of great importance. This conclusion is valid for both probabilistic and categorical metrics and, in the latter case, probability thresholds for ensembles were always observed to be lower than the average threshold among the members. When using an unconstrained linear combination, metric values are typically up to 5\% (for M-class forecasts) and 15\% (for X-class forecasts) better than ensembles using a constrained linear combination.

When looking at the top five performing ensembles in each separate skill metric used for this in-sample evaluation, three metrics seem to repeatedly appear for the M-class flares: BRIER, LCC, and NLCC\_$\rho$. It is worth noting that similar scores to those in Table~\ref{tbl:validation_metrics_m} were found in previous flare forecast validation studies. The tendency of forecasts to over-forecast was also found by \citet{Murray2017} and \citet{Sharpe2017} for the validation of MOSWOC forecasts. However, interestingly in this work the highest probability bins in the reliability diagrams of Figure~\ref{fig:verify_m} also over-forecast, while \citet{Murray2017} found under-forecasting for high probabilities. Brier scores generally also generally agree with these earlier works, although the comparison study of \citet{Barnes2016} found slightly lower values. However, it is difficult to gain any meaningful insight when inter-comparing works that used different sized data sets over different time periods, and as mentioned above these results do not correspond to those of a validation process because the metrics were calculated in-sample.

It is particularly interesting to note how well the simple equal-weights ensemble performs in this work compared to the more complex weighting schemes. While equal-weights ensembles will rarely outperform the human-edited forecasts, they have been successful in outperforming the best of the automated methods \citep{Murray2018}. These could be a helpful starting point for forecasters when issuing operational forecasts before additional information or more complex model results are obtained. However, the weighting schemes do provide a level of flexibility that simple averages cannot; they allow operational centers to tailor their forecasts depending on what measure of performance a user cares about the most (e.g., do they want to mitigate against misses or false alarms?). In this work, only a selection of metrics are highlighted based on current standards used by the community. However, the data used here are provided with open access so that readers can perform their own analysis (see Acknowledgements).

Ensemble models possess the great advantage of estimating forecast uncertainties, even in cases when none of the members have associated uncertainties. The two main sources of uncertainty for multi-model ensemble prediction are statistical and systematic; the former is quantified by the weighted standard deviation, while the latter depends (mostly) on the uncertainty of the combination weights. For both constrained and unconstrained linear combination ensembles, fractional uncertainties are observed to decrease non-linearly with increasing ensemble probability. However, the overall values of uncertainties are lower for the unconstrained linear combination ensemble cases. The lowest values of fractional uncertainty ($\sim 0.05 - 5$\% for $P \gtrsim 0.2$) are achieved by the BRIER ensemble. The main factor making the difference between constrained and unconstrained ensembles resides in the number of non-zero weights; the more members in an ensemble, the smaller the uncertainty.

The results presented in this study demonstrate that multi-model ensemble predictions of solar flares are flexible and versatile enough to be implemented and used in operational environments with metrics that satisfy user-specific needs. The evaluation of the ensemble forecasts is deferred to a future work since the intention of the present study is to illustrate how operational centers may implement an ensemble forecasting system for major solar flares using any number of members and optimization metric.

\begin{acknowledgements}
      The forecast data used here is available via Zenodo (\url{https://doi.org/10.5281/zenodo.3964552}). The analysis made use of the Python Scipy \citep{Oliphant2007} and R Verification \citep{Gilleland2015} packages. S.~A.~M. is supported by Air Force Office of Scientific Research (AFSOR) award number FA9550-19-1-7010, and previously by the Irish Research Council Postdoctoral Fellowship Programme and the AFOSR award number FA9550-17-1-039. Initial funding for J.~G.~A. was provided by the European Union Horizon 2020 research and innovation programme under grant agreement No. 640216 (FLARECAST project; \url{http://flarecast.eu}). The authors thank the anonymous reviewers for their comments and recommendations.
\end{acknowledgements}

\bibliography{swsc}

\begin{appendix}

\section{Categorical metrics definitions}\label{sec:cat_met}

Probabilistic forecasts $P$ are transformed into categorical ones by choosing a probability threshold value, $P_{\rm{th}}$ and then applying the transformation,

\begin{equation}
P_{\rm{cat}} = 
\begin{cases}
1 & \mathrm{if} \quad P \geq P_{\rm{th}} \\
0 & \mathrm{if} \quad P < P_{\rm{th}}
\end{cases}
\end{equation}

In this investigation the chosen value for $P_{\rm{th}}$ in every case, corresponds to that which maximizes the value of the metric in use. This threshold value is determined during the optimization process by constructing a metric vs $P_{\rm{th}}$ curve and finding the value that minimizes or maximizes the specific metric, depending on whether small or large values indicate better forecast performance.

\begin{table}[!th]
\centering
\caption{Contingency table for deterministic (yes/no) forecasts and event classes.}
\label{tbl:2x2_table}
\begin{tabular}{c|c|c}
& Event Observed: Yes (1) & Event Observed: No (0) \\
\hline
Event Forecast: Yes (1) & $a$ (hits) & $b$ (false alarms) \\
Event Forecast: No (0) & $c$ (misses) & $d$ (correct negatives) \\
\end{tabular}
\end{table}

A 2x2 contingency table (Table~\ref{tbl:2x2_table}) summarizes the four possible outcomes in case of deterministic forecasts ($P_{\rm{cat}}$) and events ($E$). Categorical metrics are derived from Table~\ref{tbl:metrics} following,

\begin{itemize}
\item Brier score: BRIER\_C $= \frac{1}{N}\sum (P_{\rm{cat}} - E)^{2}$
\item True Skill Score: TSS $ = \frac{ad - bc}{(a+c)(b+d)}$
\item Heidke Skill Score: HSS $ = \frac{a + d - e}{n - e}$ with $n = a + b + c + d$ and $e = (a+b)(a+c) + (b+d)(c+d)$
\item Accuracy : ACC $ = \frac{a + d}{n}$
\item Critical Success Index: CSI $ = \frac{a}{a + b + c}$
\item Gilbert Skill Score: GSS $ = \frac{a - a_{\rm{random}} }{a + b + c - a_{\rm{random}}}$ with $a_{\rm{random}} = \frac{(a + c)(a + b)}{n}$
\end{itemize}

\section{Forecast Comparison Metrics}

\begin{table}[!h]
\caption{Table with validation metrics for M-class flare forecasts. Note that there are 189 event days and 907 non-event days out of 1096 total days, and for all cases the decomposed Brier uncertainty is 0.143.}
\centering
\begin{tabular}{l l c c c c c}
\hline \hline
Ensemble        & Forecast Name /	& Brier 	& Reliability 	& Resolution 	& ROC 		& Group 	\\
Grouping        & Ensemble Label	& Score 	&  				& 				& Area		& Rank 		\\

\hline \hline
Members         & ASAP		    	& 0.151		& 0.0163 		& 0.0079		& 0.575		& 7			\\
                & ASSA		    	& 0.150		& 0.0235 		& 0.0167		& 0.738		& 5			\\
                & MAG4		    	& 0.126		& 0.0064 		& 0.0237		& 0.772		& 4			\\
                & MCSTAT			& 0.183		& 0.0606 		& 0.0200		& 0.769		& 6			\\
                & MOSWOC			& 0.116		& 0.0056 		& 0.0327		& 0.842		& 1			\\
                & NOAA		    	& 0.116		& 0.0070 		& 0.0335		& 0.838		& 2			\\
\cline{2-7}
                & Equal-weights 	& 0.121		& 0.0046 		& 0.0264		& 0.816		& 3			\\
\hline
Prob.-optimized & BRIER	 	    	& 0.110		& 0.0009 		& 0.0338		& 0.848		& 8			\\
                & BRIER\_unc		& 0.107		& 0.0007 		& 0.0368		& 0.853		& 2			\\
                & LCC				& 0.109		& 0.0016 		& 0.0355		& 0.848		& 6			\\
                & LCC\_unc	    	& 0.106		& 0.0019 		& 0.0387		& 0.853		& 2			\\
                & MAE				& 0.126		& 0.0064 		& 0.0237		& 0.772		& 15		\\
                & MAE\_unc	    	& 0.127		& 0.0082 		& 0.0244		& 0.811		& 15		\\
                & NLCC\_$\rho$	    & 0.110		& 0.0011 		& 0.0334		& 0.849		& 7			\\
                & NLCC\_$\rho$\_unc	& 0.107		& 0.0007 		& 0.0366		& 0.854		& 1			\\
                & NLCC\_$\tau$	    & 0.110		& 0.0018 		& 0.0344		& 0.848		& 8			\\
                & NLCC\_$\tau$\_unc	& 0.109		& 0.0011 		& 0.0351		& 0.856		& 2			\\
                & REL				& 0.114		& 0.0013 		& 0.0298		& 0.831		& 14		\\
                & REL\_unc	    	& 0.111		& 0.0008 		& 0.0322		& 0.841		& 12		\\
                & RES				& 0.110		& 0.0009 		& 0.0332		& 0.841		& 11		\\
                & RES\_unc	    	& 0.114		& 0.0010 		& 0.0322		& 0.832		& 13		\\
                & ROC				& 0.109		& 0.0021 		& 0.0357		& 0.847		& 8			\\
                & ROC\_unc	    	& 0.108		& 0.0010 		& 0.0357		& 0.853		& 5			\\
\hline
Cat.-optimized  & ACC	        	& 0.112		& 0.0023 		& 0.0335		& 0.890		& 10		\\
                & ACC\_unc	        & 0.126		& 0.0131 		& 0.0297		& 0.625		& 11		\\
                & BRIER\_C		    & 0.111		& 0.0008 		& 0.0327		& 0.891		& 8			\\
                & BRIER\_C\_unc     & 0.129		& 0.0156 		& 0.0289		& 0.596		& 12		\\
                & CSI				& 0.109		& 0.0013 		& 0.0350		& 0.889		& 1			\\
                & CSI\_unc	    	& 0.111		& 0.0062 		& 0.0376		& 0.630		& 2			\\
                & GSS				& 0.110		& 0.0008 		& 0.0338		& 0.839		& 3			\\
                & GSS\_unc	    	& 0.129		& 0.0221 		& 0.0360		& 0.878		& 7			\\
                & HSS				& 0.111		& 0.0033 		& 0.0349		& 0.889		& 7			\\
                & HSS\_unc	    	& 0.108		& 0.0021 		& 0.0372		& 0.620		& 5			\\
                & TSS				& 0.111		& 0.0013 		& 0.0348		& 0.856		& 3			\\
                & TSS\_unc	    	& 0.130		& 0.0227 		& 0.0359		& 0.879		& 9			\\
\hline
\end{tabular}
\label{tbl:validation_metrics_m}
\end{table}

\begin{table}[!h]
\caption{Table with validation metrics for X-class flare forecasts. Note that there are 17 event days and 1,079 non-event days out of 1096 total days, and for all cases the decomposed Brier uncertainty is 0.015.}
\centering
\begin{tabular}{l l c c c c c}
\hline \hline
Ensemble        & Forecast Name /	& Brier 	& Reliability 	& Resolution 	& ROC 		& Group 	\\
Grouping        & Ensemble Label	& Score 	&  				& 				& Area		& Rank 		\\

\hline \hline
Members         & ASAP		    	& 0.047		& 0.0319 		& 0.0002		& 0.534		& 7			\\
                & ASSA		    	& 0.018		& 0.0028 		& 0.0000		& 0.716		& 6			\\
                & MAG4		    	& 0.016		& 0.0030 		& 0.0024		& 0.767		& 2			\\
                & MCSTAT			& 0.026		& 0.0115 		& 0.0008		& 0.878		& 3			\\
                & MOSWOC			& 0.017		& 0.0049 		& 0.0035		& 0.879		& 1			\\
                & NOAA		    	& 0.018		& 0.0046 		& 0.0015		& 0.834		& 3			\\
\cline{3-7} 
                & Equal-weights	    & 0.019		& 0.0045 		& 0.0008		& 0.874		& 3			\\
\hline
Prob.-optimized & BRIER		    	& 0.016		& 0.0030		& 0.0027		& 0.888		& 4			\\
                & BRIER\_unc		& 0.015		& 0.0023 		& 0.0023		& 0.820		& 7			\\
                & LCC				& 0.016		& 0.0030 		& 0.0027		& 0.896		& 1			\\
                & LCC\_unc	    	& 0.015		& 0.0031 		& 0.0033		& 0.879		& 1			\\
                & MAE				& 0.015		& 0.0024 		& 0.0024		& 0.768		& 13		\\
                & MAE\_unc	    	& 0.016		& 0.0020 		& 0.0015		& 0.680		& 15		\\
                & NLCC\_$\rho$	    & 0.018		& 0.0051 		& 0.0026		& 0.909		& 4			\\
                & NLCC\_$\rho$\_unc	& 0.016		& 0.0028 		& 0.0017		& 0.906		& 3			\\
                & NLCC\_$\tau$	    & 0.018		& 0.0050 		& 0.0023		& 0.909		& 4			\\
                & NLCC\_$\tau$\_unc	& 0.020		& 0.0067 		& 0.0018		& 0.919		& 7			\\
                & REL				& 0.017		& 0.0027 		& 0.0014		& 0.896		& 7			\\
                & REL\_unc	    	& 0.015		& 0.0019 		& 0.0018		& 0.865		& 10		\\
                & RES				& 0.017		& 0.0040 		& 0.0027		& 0.895		& 14		\\
                & RES\_unc	    	& 0.016		& 0.0024 		& 0.0014		& 0.894		& 10		\\
                & ROC				& 0.018		& 0.0052 		& 0.0025		& 0.908		& 10		\\
                & ROC\_unc	    	& 0.026		& 0.0132 		& 0.0021		& 0.887		& 16		\\
\hline
Cat.-optimized  & ACC	    	    & 0.017 	& 0.0033		& 0.0018		& 0.890		& 2			\\
                & ACC\_unc	        & 0.016 	& 0.0032		& 0.0027		& 0.625		& 7			\\
                & BRIER\_C		    & 0.017 	& 0.0033		& 0.0017		& 0.891		& 2			\\
                & BRIER\_C\_unc     & 0.016 	& 0.0032		& 0.0027		& 0.596		& 9			\\
                & CSI				& 0.016		& 0.0030		& 0.0026		& 0.889		& 1			\\
                & CSI\_unc	    	& 0.015 	& 0.0038		& 0.0045		& 0.630		& 5			\\
                & GSS				& 0.022 	& 0.0076		& 0.0009		& 0.839		& 12		\\
                & GSS\_unc	    	& 0.019 	& 0.0055		& 0.0021		& 0.878		& 9			\\
                & HSS				& 0.016 	& 0.0034		& 0.0030		& 0.889		& 2			\\
                & HSS\_unc	    	& 0.015 	& 0.0038		& 0.0046		& 0.620		& 6			\\
                & TSS				& 0.021 	& 0.0068		& 0.0010		& 0.856		& 11		\\
                & TSS\_unc	    	& 0.018 	& 0.0050		& 0.0024		& 0.879		& 7			\\
\hline
\hline
\end{tabular}
\label{tbl:validation_metrics_x}
\end{table}

\section{X-class Flare Forecast Results}\label{sec:app_xclass}

This section contains results similar to those presented in Sections \ref{sec:comb_w} and \ref{sec:opt_metrics} for the case of X-class flare forecasts.
\begin{figure}[h]
\centering

\includegraphics[width=34pc]{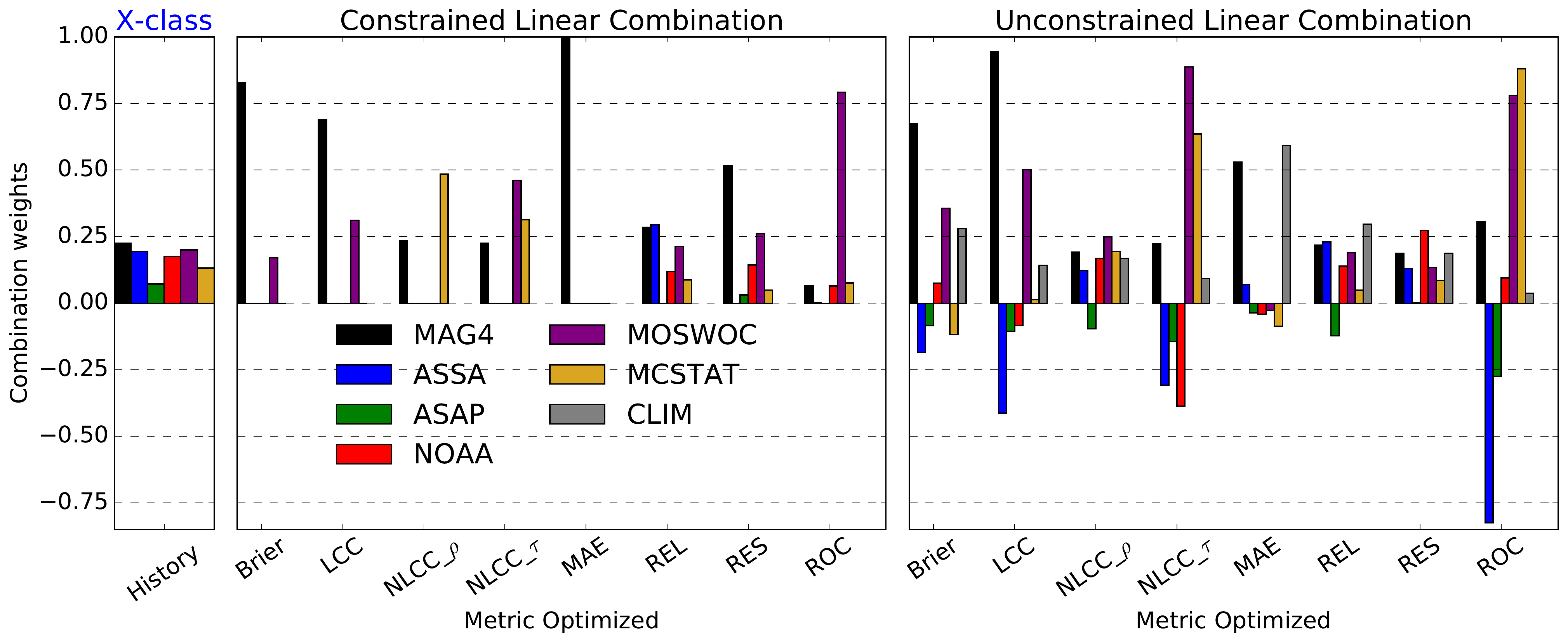}
\caption{Same as Figure~\ref{fig:w_prob_m}, but for X-class flare forecasts.}
\label{fig:w_prob_x}
\end{figure}

\begin{figure}[h]
\centering

\includegraphics[width=34pc]{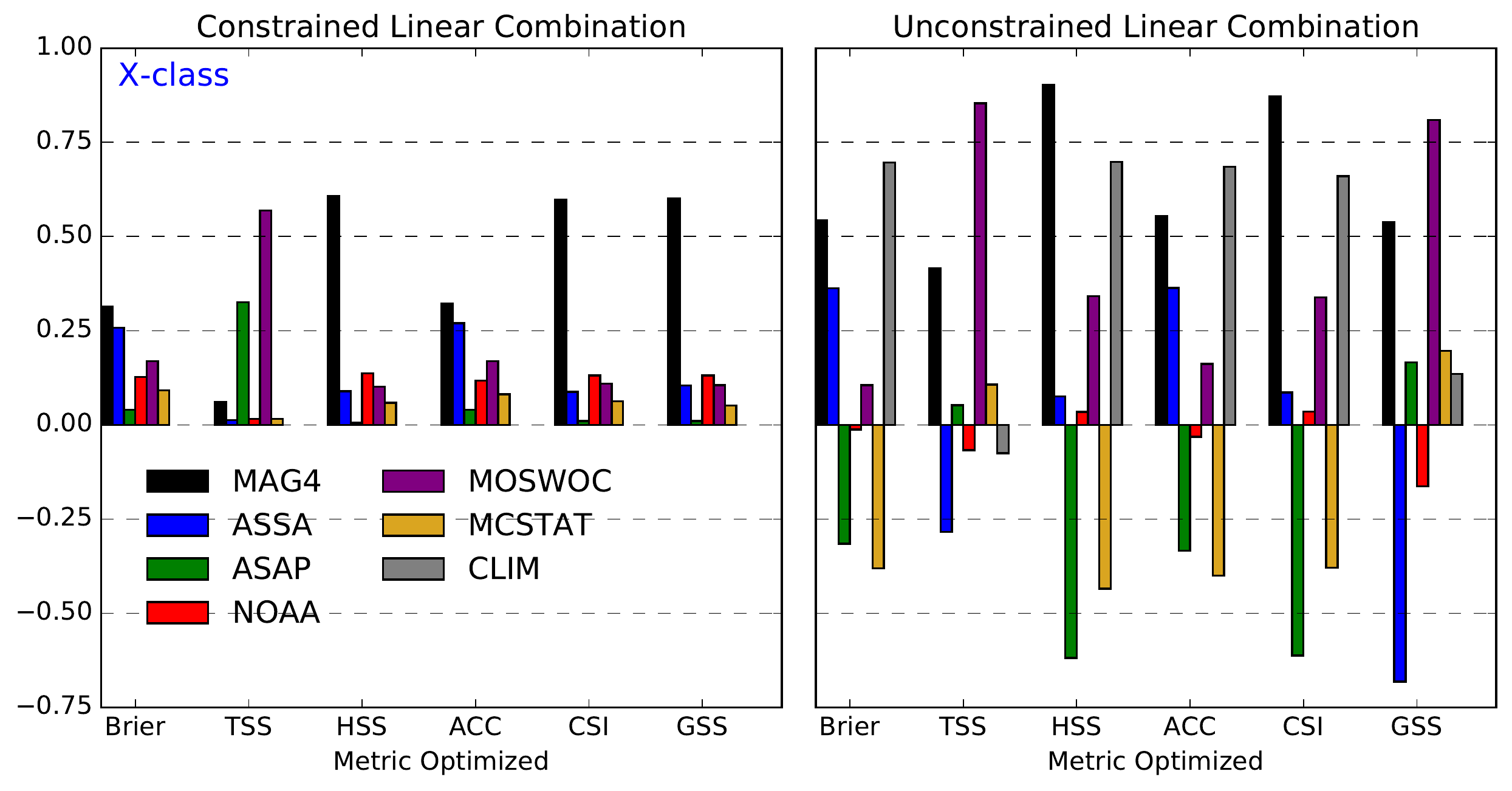}

\caption{Same as Figure~\ref{fig:w_cat_m}, but for X-class flare forecasts.}
\label{fig:w_cat_x}
\end{figure}

\begin{figure}[h]
\centering

\includegraphics[width=18pc]{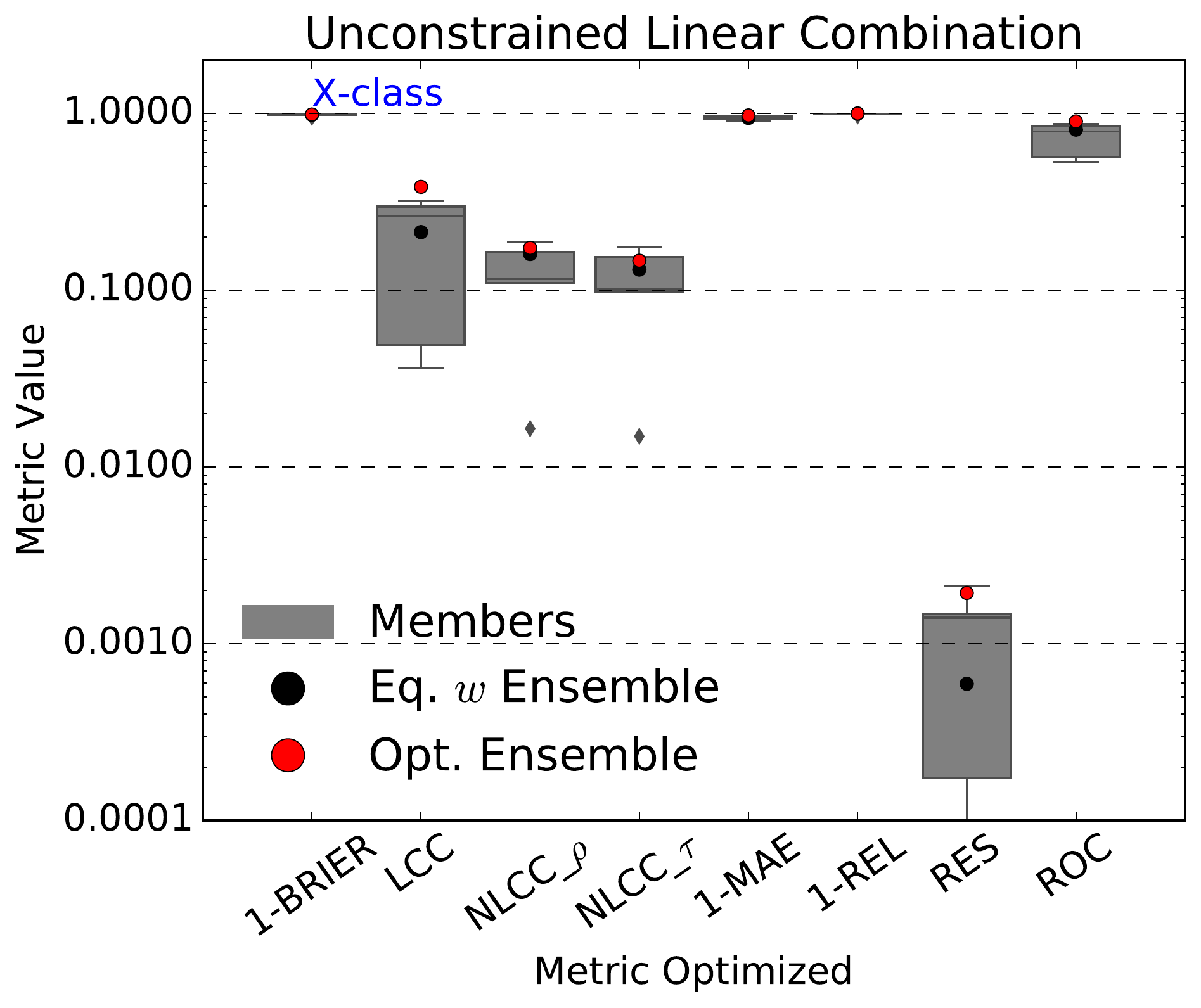}
\includegraphics[width=18pc]{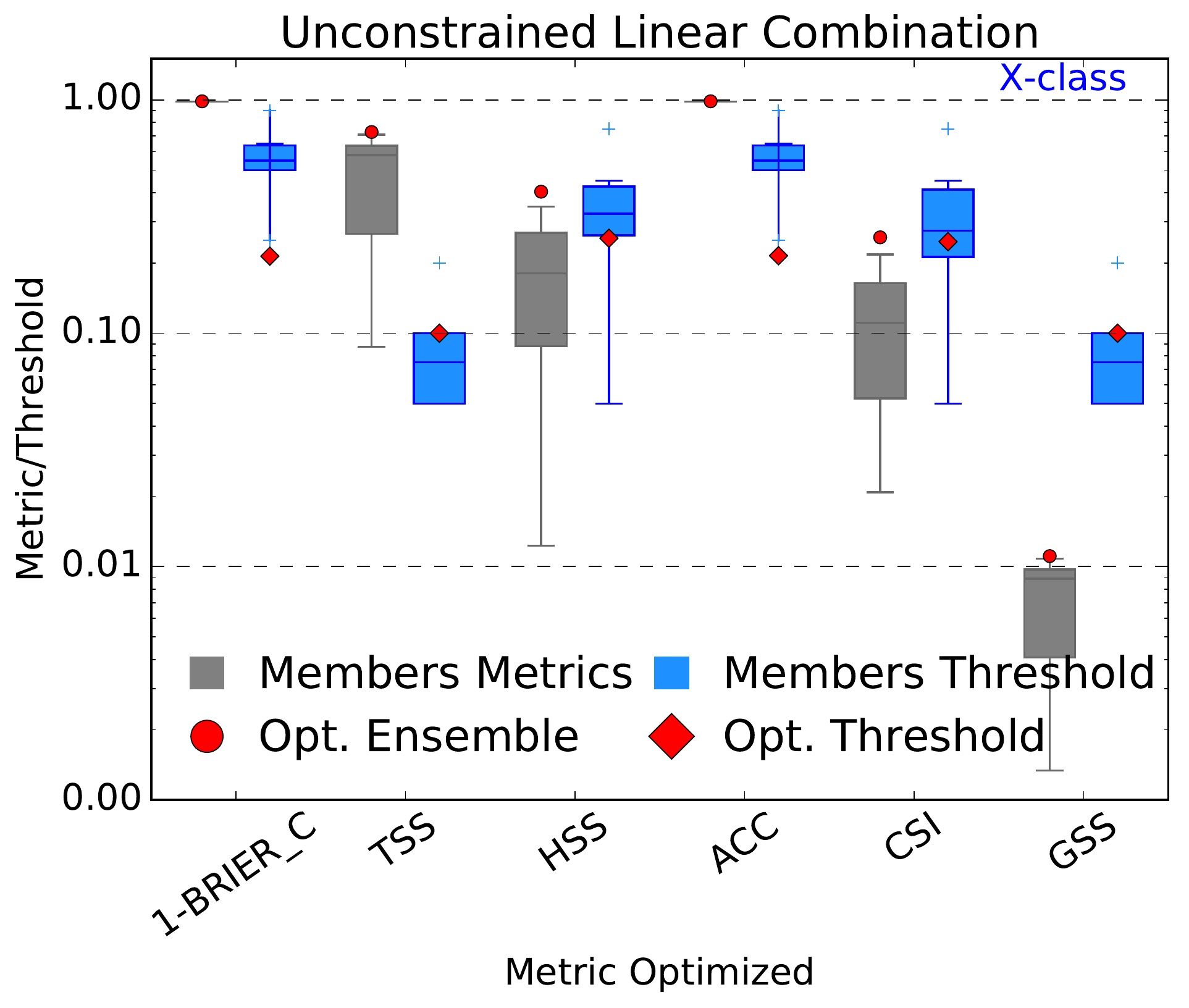}
\caption{Same as Figure~\ref{fig:metrics_m_unconst}, but for X-class flare forecasts.}
\label{fig:metrics_cat_x}
\end{figure}

\begin{figure}[!t]
\centering
\includegraphics[width=\textwidth]{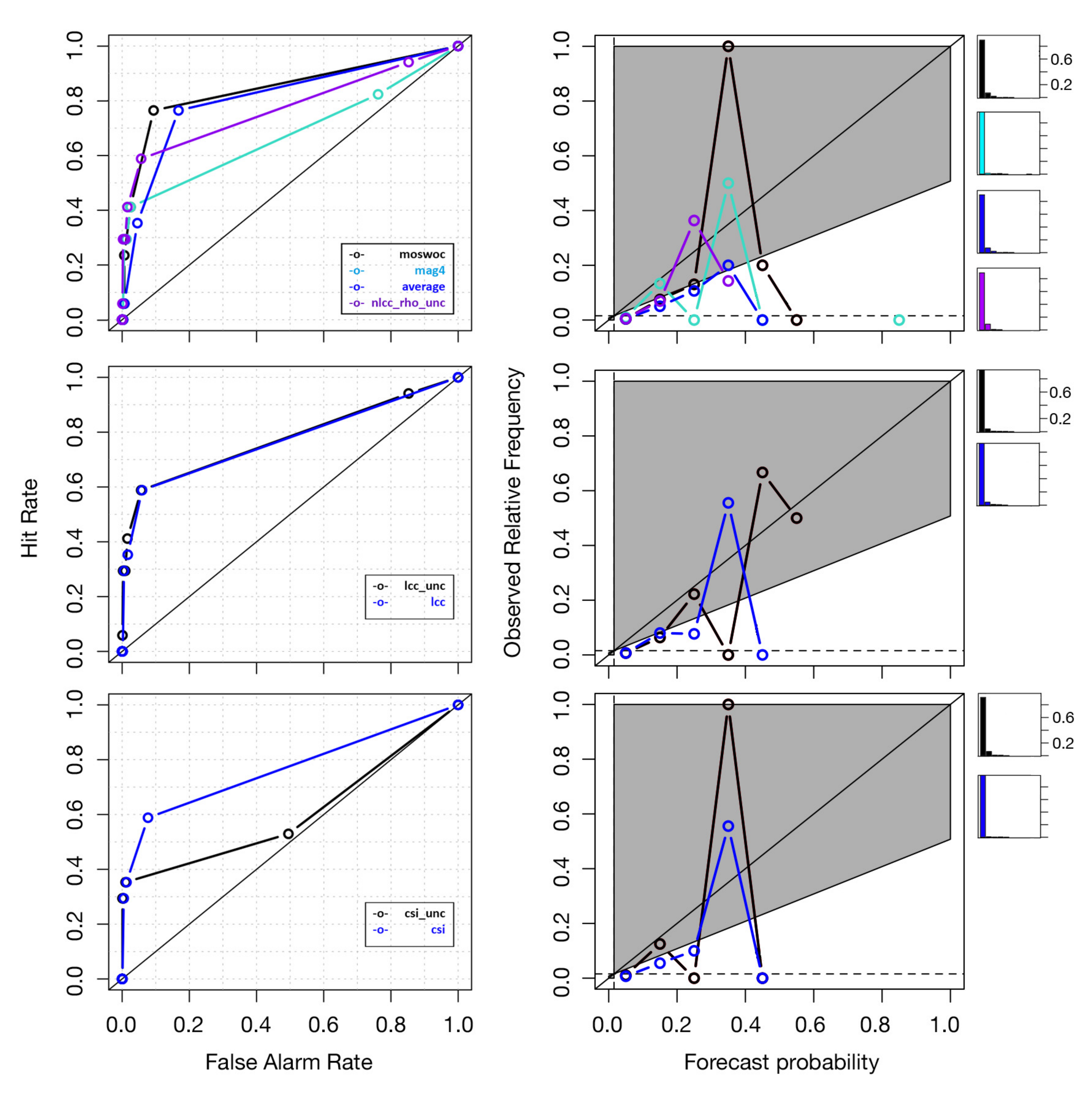}
\caption{Same as Figure~\ref{fig:verify_m}, but for X-class flare forecasts.}
\label{fig:verify_x}
\end{figure}

\begin{table}[!t]
\caption{Rankings of evaluation metrics for X-class flare forecasts. For each metric,the top five performing ensembles are displayed.}
\centering
\begin{tabular}{l c c c c}
\hline \hline
Rank 	& Brier					& Reliability 		& Resolution 	& ROC 				\\
		& score					&  					& 				& area				\\
\hline \hline
1		& HSS\_unc				& REL\_unc			& HSS\_unc		& NLCC\_$\tau$\_unc  	\\
2		& CSI\_unc  			& MAE\_unc 			& CSI\_unc		& NLCC\_$\tau$			\\
3		& LCC\_unc      		& BRIER\_unc 	    & MOSWOC		& NLCC\_$\rho$			\\
4		& BRIER\_unc	    	& MAE 				& LCC\_unc		& ROC				\\
5		& MAE\_unc				& RES\_unc 			& HSS			& NLCC\_$\rho$\_unc	\\	
\hline
\end{tabular}
\label{tbl:validation_metric_ranks_x}
\end{table}

\end{appendix}

\end{document}